\documentclass[twocolumn,showpacs,preprintnumbers,amsmath,amssymb]{revtex4}
\usepackage{graphicx}
\usepackage{bm}
\usepackage{epsf}
\usepackage{xcolor}
\usepackage{epstopdf}
\begin{document}
\title{Sublimation of silicene and thin silicon films: a view from molecular dynamics simulation}

\author{Yu. D. Fomin \footnote{Corresponding author: fomin314@mail.ru}}
\affiliation{Vereshchagin Institute of High Pressure Physics,
Russian Academy of Sciences, Kaluzhskoe shosse, 14, Troitsk,
Moscow, 108840, Russia }

\author{E. N. Tsiok}
\affiliation{Vereshchagin Institute of High Pressure Physics,
Russian Academy of Sciences, Kaluzhskoe shosse, 14, Troitsk,
Moscow, 108840, Russia }

\author{V. N. Ryzhov }
\affiliation{Vereshchagin Institute of High Pressure Physics,
Russian Academy of Sciences, Kaluzhskoe shosse, 14, Troitsk,
Moscow, 108840, Russia }
\date{\today}

\begin{abstract}

A molecular dynamics simulation of sublimation of silicene and silicon films of different thikness is performed.
It is shown that thiner films sublimate at lower temperatures. The sublimation temperature comes to a saturated
value of $T=1725$ K at the films thiker than $16$ atomc layers. These results are consistent with the surface mediated
collaps of the crystal structure. At the same time this mechanism is different from the crystal structure collapse of graphite
and graphene.

\end{abstract}


\maketitle


\section{Introduction}

Nowadays there is a great interest to low-dimensional materials such as graphene. It is expected that such materials might
strongly improve many modern technologies. One of the principle technologies is electronics, which in currently based on
silicon. As it was predicted in Ref. \cite{si-1} as early as 1994, silicon can form a quasi-two-dimensional (q-2d) structure, called silicene.
Silicene was synthesized in 2012 \cite{si-syn} and nowadays it is expected that in can be used in different technologies,
including electronics \cite{si-electronics}, hydrogen storage \cite{si-h}, Li-ion batteries \cite{si-li-ion} and construction
of different devices \cite{si-d1,si-d2}.

A lot of experimental efforts were paid to investigation of different properties of silicene, which are summurized in several
recent reviews \cite{rev-1,rev-2}. At the same time, since the first prediction of stability of one-layer Si system \cite{si-1}, numerous
DFT calculations of silicene were reported (see, i.e. \cite{dft-1,dft-2,dft-3,dft-4,dft-5,dft-6}, to name a few). Some problems are more
appropriate to study within the framework of classical molecular dynamics or Monte-Carlo simulations with empirical potentials. Such problems are
related to the cases when we need to simulate large systems (at least thousands of atoms), long time intervals (at least nanoseconds) or high
temperatures.

A particular point of interest is the melting temperature of silicene. The melting temperature of bulk silicon is $1683$ K \cite{tm}. One should expect that
the silicene melting point should be at least as of the same order of magnitude, which makes experimental measurements to be difficult.
For this reason the melting point of silicene was probed by classical molecular dynamics and Monte Carlo simulation in a number of works.
In Ref. \cite{melting-1} thermal stability of silicene was simulated. Tersoff potential \cite{tersoff} was employed with two different sets of
parameters: the original one by Tersoff \cite{tersoff} and the one from Ref. \cite{tersoff-a}. The melting temperature of silicene is found to be
$T_m=3600$ K in the case of original parameters and $T_m=1750$ K for the ones from Ref. \cite{tersoff-a}. Strong dependence of the melting
point on the details of employed model is evident.

Tefsoff potential was also used in the work \cite{melting-2}, but a non-equilibrium molecular dynamics method was employed in this work, rather than
Monte Carlo. The melting  point reported in \cite{melting-2} is $T_m=1201$ K, which is much lower than the experimental melting point of bulk silicon
and the one obtained in Ref. \cite{melting-1}.

Non-equilibrium molecular dynamics simulation was performed in Ref. \cite{melting-3}, but a Stillinger-Weber (SW) potential \cite{sw} was employed in this paper.
A sample of silicene was heated up with the heating rate $\frac{dT}{dt}=1.7 \cdot 10^{12}$ {K/s}. The melting point reported in this paper
is $T_m=1500$ K. The same melting point was obtained in Ref. \cite{melting-4} within ReaxFF model \cite{reax}.

The same SW potential was used in Ref. \cite{melting-5}. The general methodology of this work is similar to the one of Ref. \cite{melting-3}, but
the heating rate is $\frac{dT}{dt}=10^{13}$ $K/s$. The melting temperature obtained in this work is $T_m=2500$ K.

From this discussion one can see that the melting point obtained in simulation strongly depends on employed model and the simulation setup and ranges
from $T_{low}=1201$ K up to $T_{high}=3600$ K.

In our recent paper \cite{carbon} "melting" of graphene was discussed. Most of the publications on graphite melting employ the same
non-equilibrium methodology that is used for silicene, i.e. a sample of graphene is heated with some heating rate in molecular
dynamics simulation until the crystal structure collapses. We have shown that  indeed sublimation rather then melting is observed in such simulations.

The same reasoning which was used for "melting of graphene" \cite{carbon} can be employed for silicene too: melting is possible only if the
pressure exceeds the one in the triple point. At the same time simulation of a layer of silicene surrounded by vacuum
implies zero pressure, therefore sublimation should take place. It might be seen from the literature data: Figures 4 and 6 of Ref.  \cite{melting-3}
shows snapshots of silicene after "melting" where one can observe that the system splits into cylinders or balls. This phenomenon is observed
in simulation of liquid - vapor two phase region and related to finite size effects of simulation \cite{prest}.

In the present paper we study the dependence of the temperature of collapse of the crystal structure of silicon on the width of the sample. We start from
silicene layer (one layer system), then we study the system with diamond structure with four layers, eight layers, etc. up to twenty ones and compare the results
with the melting point of the bulk system.

\section{System and Methods}

In the present work we investigate crystal structure collapse of a silicon sample of different width by means of molecular dynamics simulation. Stillinger-Weber
potential for silicon is used \cite{sw}. The system consisted of a silicon sample surrounded by vacuum.  Periodic boundary conditions were employed in
all three directions. In order to minimize the influence of periodic boundaries on the results the length of the sample along z axis (perpendicular to the sample)
was choosen to be very large - $543$ $\AA$, which corresponds to a hundred of lattice constants of silicon.
The unit cell parameters of silicene are $a=b=3.861$ $\AA$, and $\alpha=\beta=90^o$ and $\gamma=120^o$.
It leads to a triclinic simulation box. In the case of diamond structure with different number of layers a box of square shape in XY plane is employed. The number of
atoms in all considered systems is given in the Table I.

\begin{table}
\begin{tabular}{ l c r }
  System & Number of particles \\
\hline
  silicene & 20000  \\
\hline
  8 layers  & 4096 \\
\hline
 12 layers & 6144 \\
\hline
 16 layers & 8192 \\
\hline
 20 layers & 10240 \\
\hline
 bulk  & 10240 \\
\hline
\end{tabular}
\caption{Number of atoms in systems under consideration.}
\end{table}

In all cases we start from ideal crystal structure and simulate the system in canonical ensembe (constant number of particles N, volume V and temperature T) at given temperature.
The set of temperatures depends on the system. The goal of this work is to find out at which temperature the crystalline structure collapses. The time step
is 1 fs and all simulations consist of $1 \cdot 10^7$ steps, i.e. 10 ps.

During the coarse of simulation we monitor the energy per particle in the system, the width of the sample (not applicable to the bulk sample). The structure of a
sample was probed by Bond Orientational Order (BOO) parameters \cite{boo}:

\begin{equation}
q_l= \frac{4 \pi}{2l+1} \sum_{m=-l}^l Q_{lm}^2,
\end{equation}
where

\begin{equation}
  Q_{lm}({\bf r})=\frac{1}{NN} \sum_{j=1}^{NN} Y_{lm}(\theta({\bf r}), \phi({\bf r})),
\end{equation}
where the summation is taken over the nearest neighbors of a given particle, $NN$ is the number of nearest neighbors, $Y_{lm}$ are spherical harmonics based on the
angles to the nearest neighbor. Parameter $q_3$ is used for the diamond structure. For ideal diamond crystal the value of the BOO is $q_3^{id}=0.745356$.
Following the work \cite{ljconf}, we divide all particles into diamond-like and disorderd. A particle is called diamond-like if $|q_3(i)-q_3^{id}|<0.1$, otherwise it is
disordered. We also calculate the probability distribution of the BOO $q_3$ in order to see its the most probable values.

All simulations were performed with lammps simulation package \cite{lammps}.

\section{Results and Discussion}

\subsection{Silicene - one layers system}

\begin{figure}

\includegraphics[width=8cm]{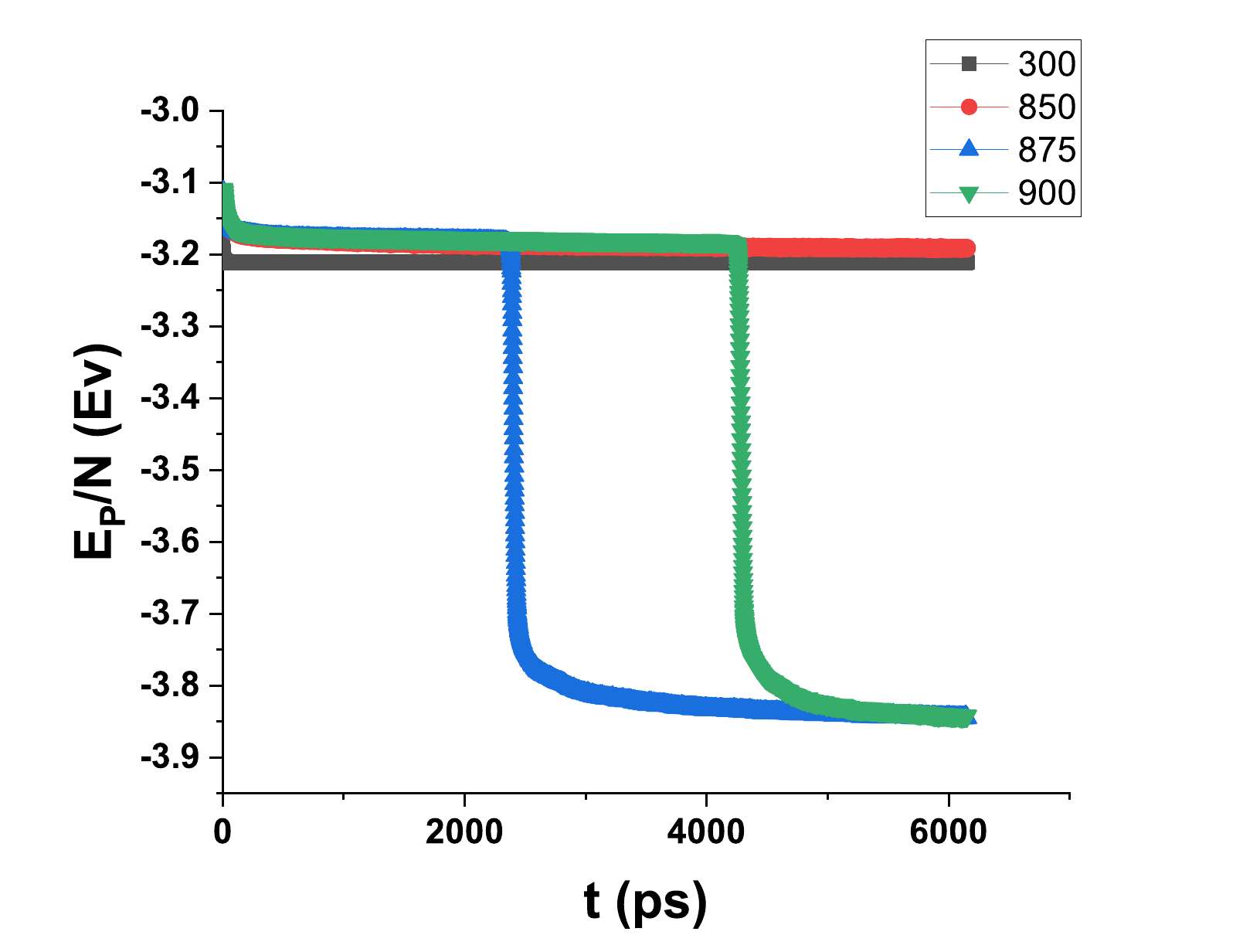}%

\caption{\label{esil} The time dependence of potential energy per particle of silicene at different temperatures.}
\end{figure}

We start the investigation from simulation of silicene at different temperatures. Figure \ref{esil} shows a time dependence of the potential energy per particle
of silicene at different temperatures. It is seen from the figure that the energy does not demonstrate any drastic changes for the temperatures below $850$ K.
However, as soon as the temperature reaches $T=875$ K it experiences an abrupt drop after some time, which means that the structure of the system has changed.

\begin{figure}

\includegraphics[width=8cm]{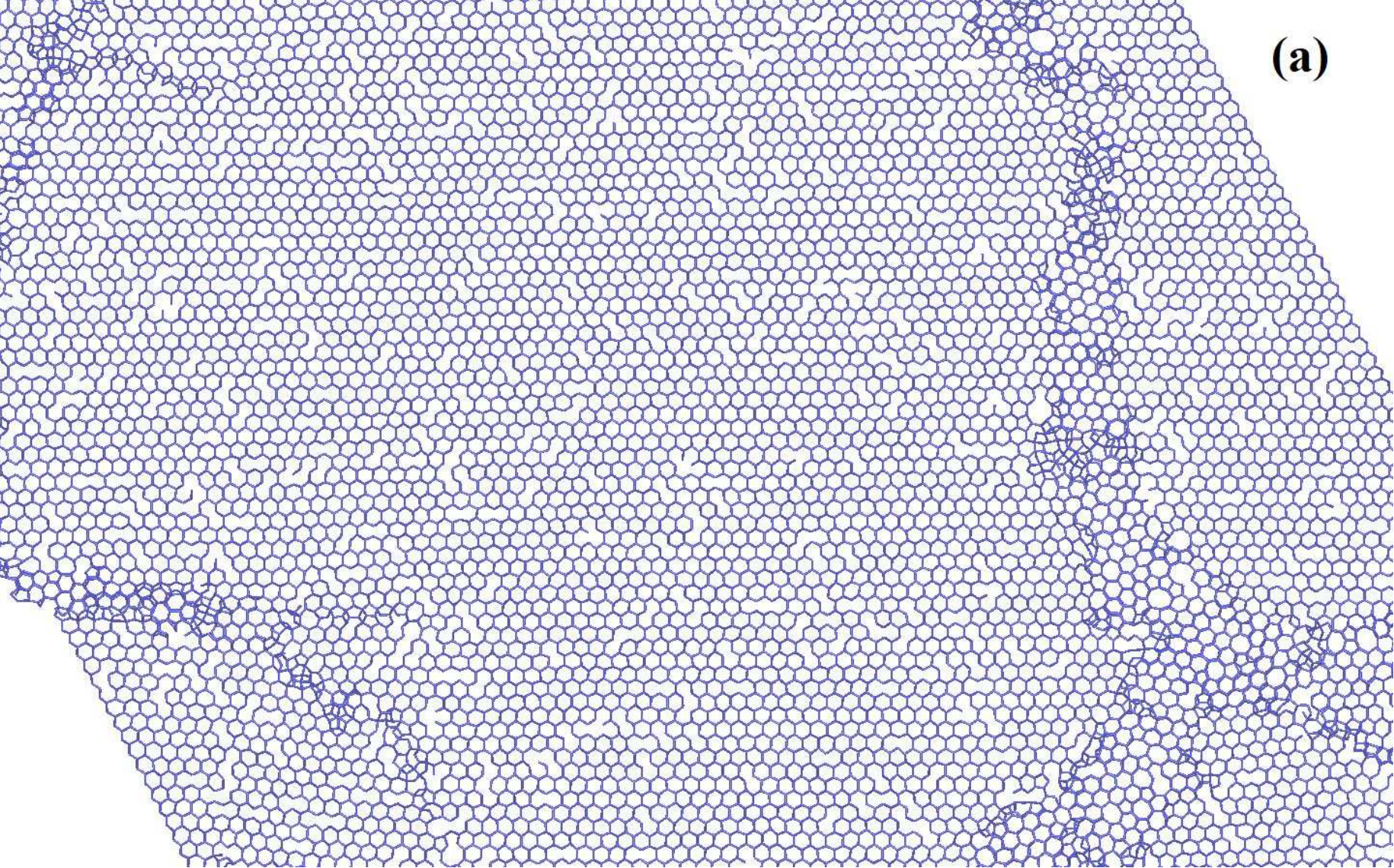}%

\includegraphics[width=8cm]{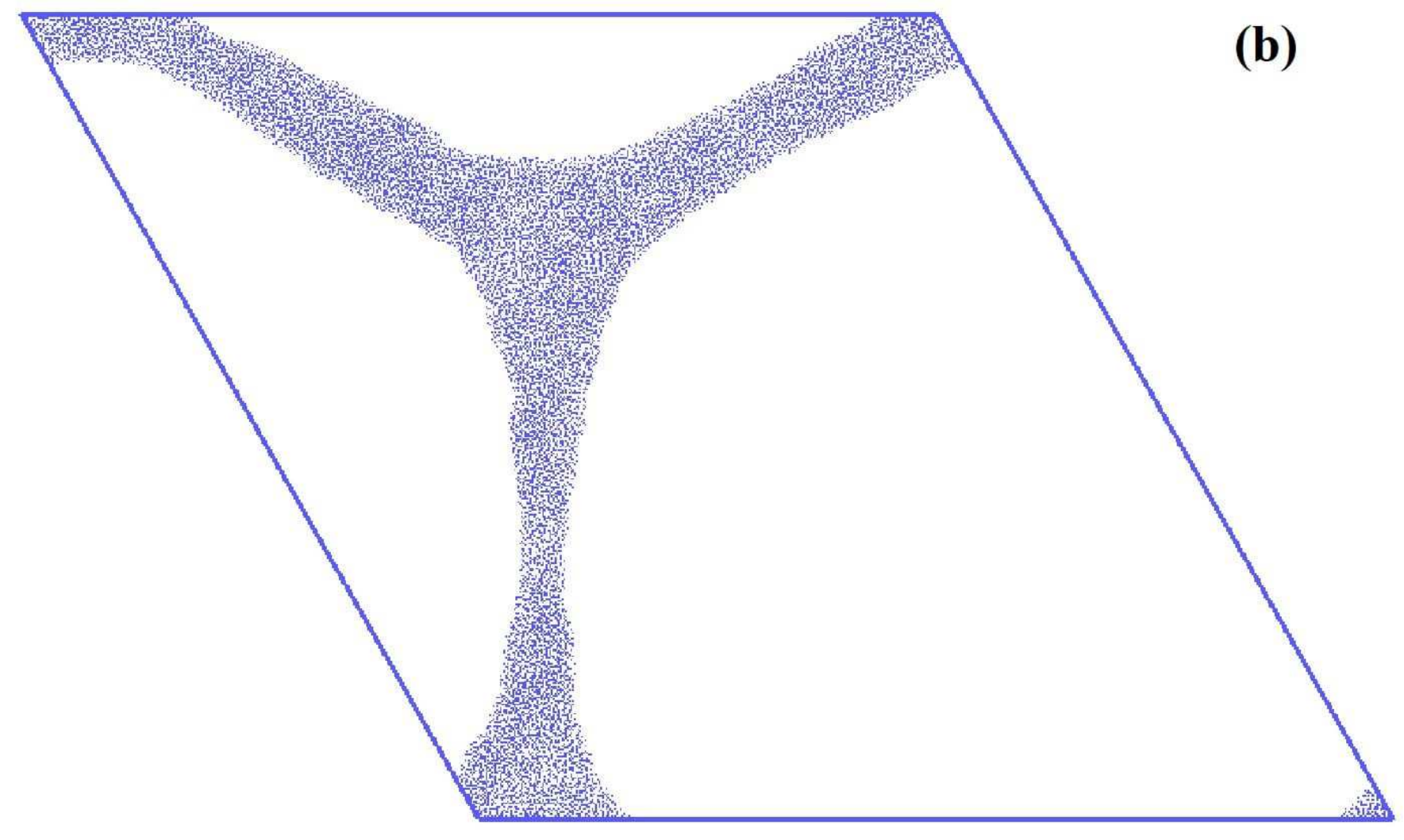}%

\caption{\label{sil-conf} The snapshots of a final configuration of the system with initial state consisting of silicene surrounded by vacuum. (a) $T=850$ K; (b) $T=875$ K.}
\end{figure}

\begin{figure}

\includegraphics[width=8cm]{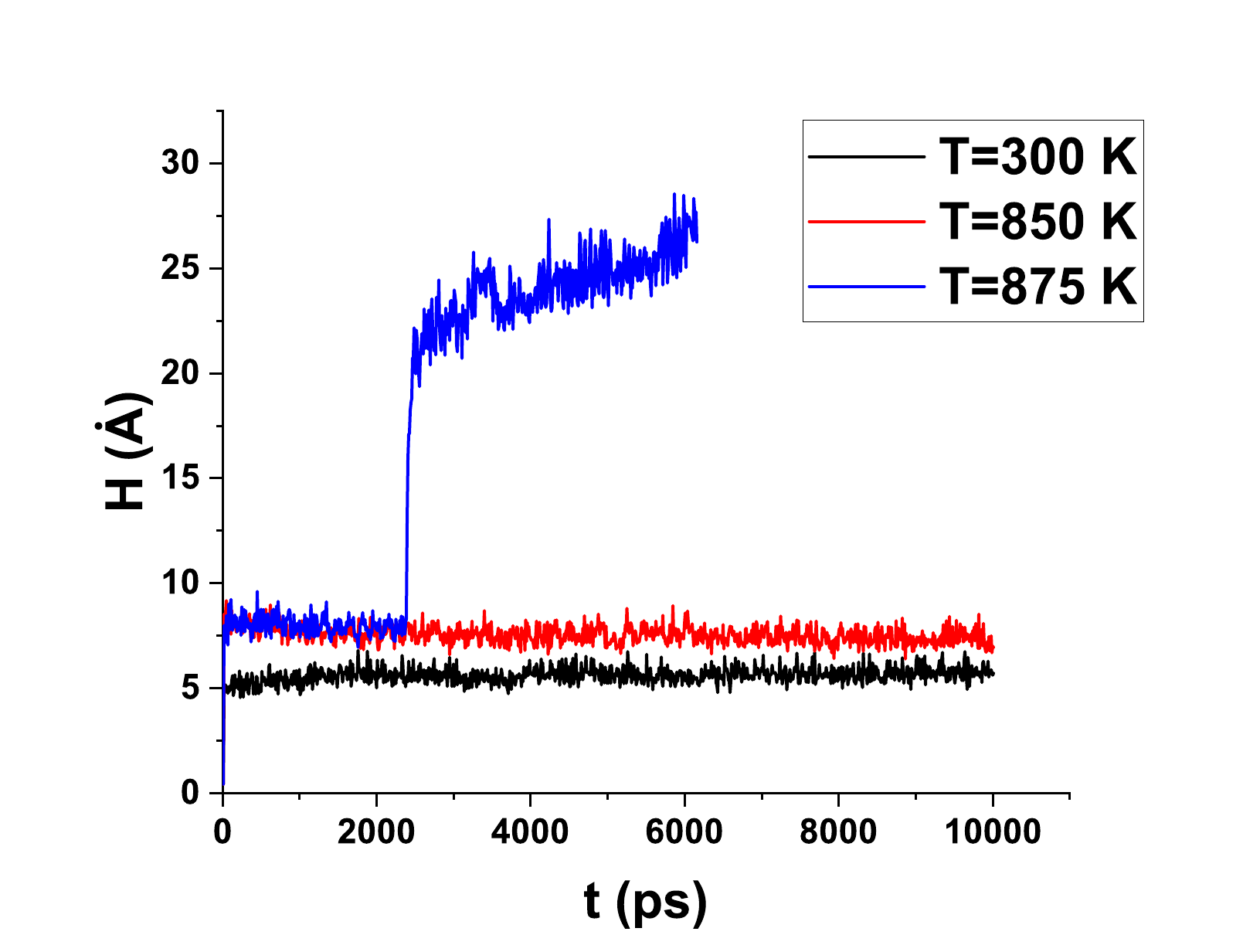}%

\caption{\label{wsil} Time dependence of the width of the sample of silicene at different temperatures. }
\end{figure}

The conclusion above is illustrated in Fig. \ref{sil-conf} where final structures of the systems at $T=850$ and $875$ K are shown. As is seen from these snapshots,
in the case of $T=850$ K the system remains in the state of strongly defected silicene layer in vacuum. At the same time at $T=875$ K the crystalline structure collapses,
and the system demonstrates a kind of disordered state.  This phase is still condensed, since it does not occupy the whole volume of the system. However, the pressure
of the system is nearly zero. We are not aware of any data on the location of gas-liquid-crystal triple point of silicon, however, it must take place at positive pressure.
Therefore, this phase should transform into vapor, which is not observed due to the finite time of the simulation. Figure \ref{wsil} shows the width of the system as
a function of time. We see from this plot that which the width of crystalline samples fluctuates about some average value, the width of the disordered system experiences
a sudden jump at the sublimation pressure and continuously increases after that. It can be concluded from this figure that the disordered system has not reached the equilibrium
in the time of our simulation.

\subsection{Eight layers system}

\begin{figure}
\includegraphics[width=8cm]{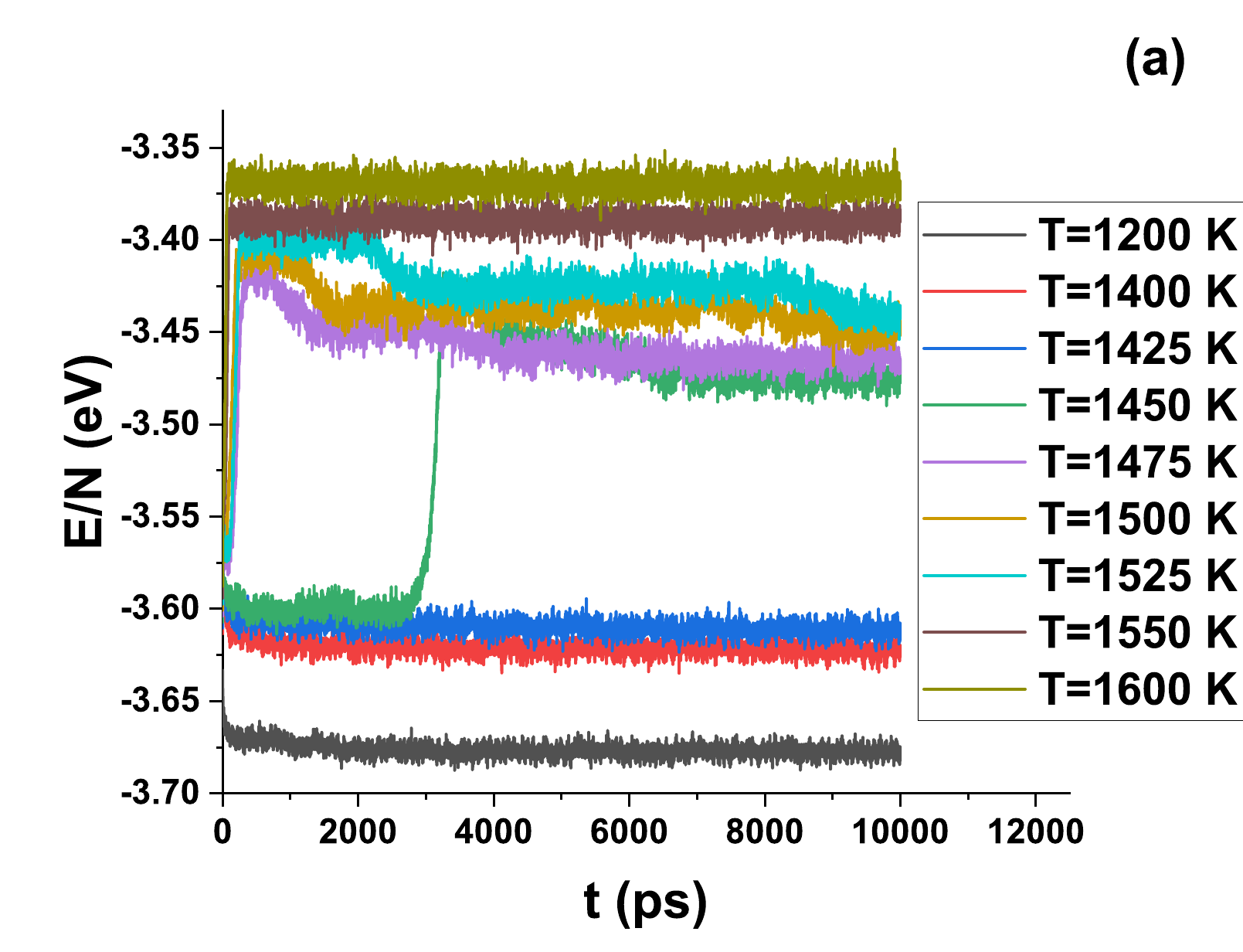}%

\includegraphics[width=8cm]{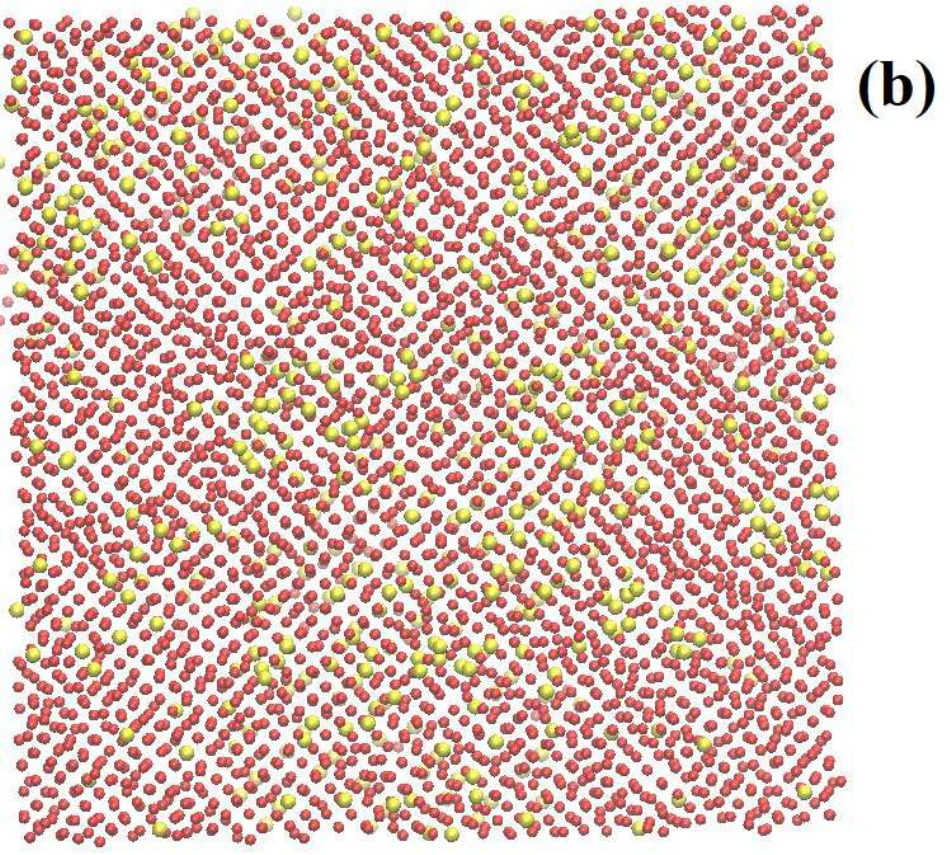}%

\includegraphics[width=8cm]{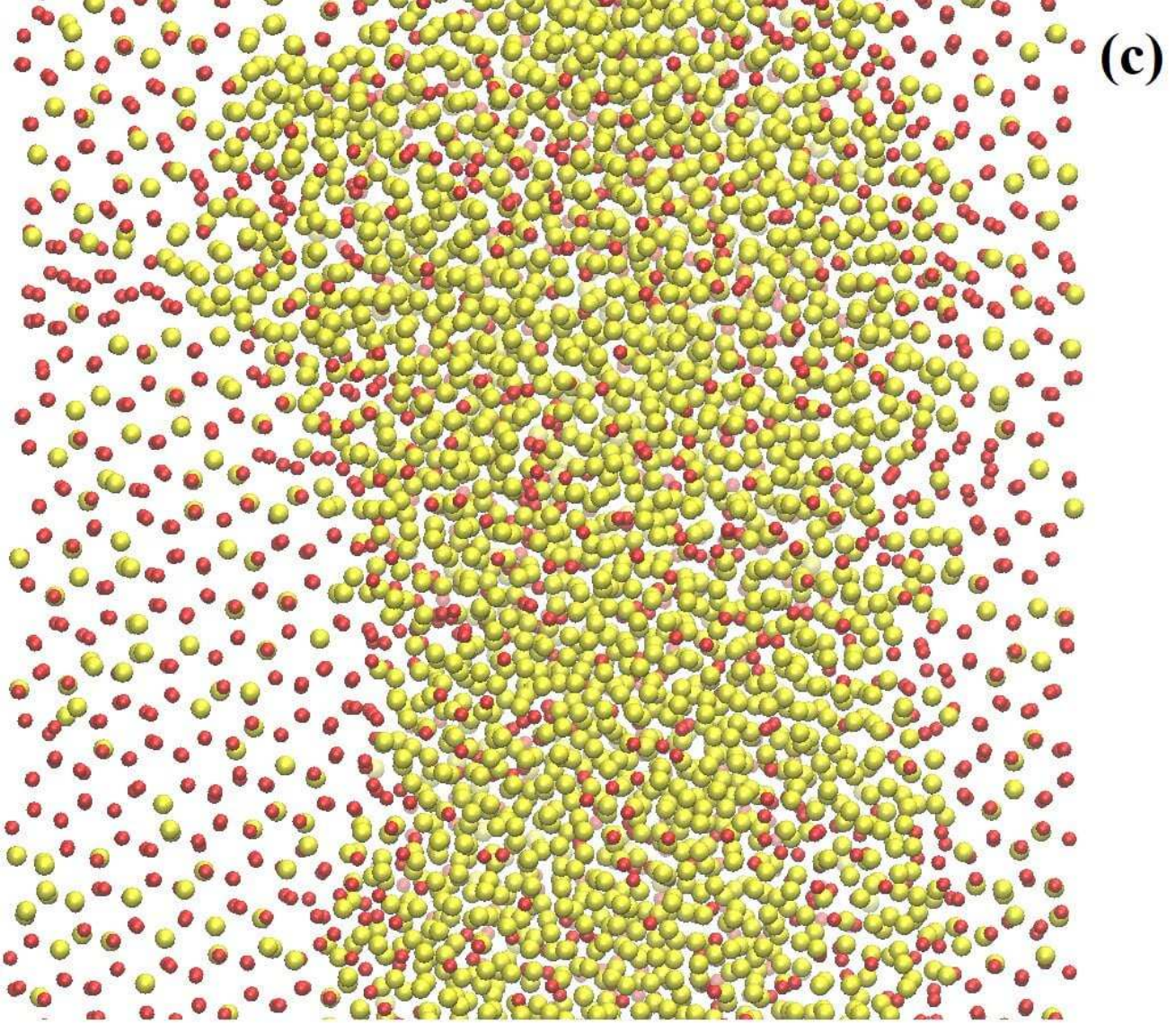}%

\caption{\label{si2} (a) Time dependence of the energy per particle of eight layer system at different temperatures. (b) A snapshot of the system at $T=1425$ K. (c) A snapshot
of the system at $T=1500$ K. Red balls are diamondlike particles, which yellow ones are the particles with disordered local surrounding (see the methons section).}
\end{figure}

We show the time evolution of energy per particle in the system with eight layers in Fig. \ref{si2} (a). As it is seen from this figure,
the energy remains constant up to the temperature $T=1425$ K, but experiences a sudden jump at $T=1450$ K. We conclude that
a change in structure of the system take place at the later temperature. Figures \ref{si2} (b) and (c) show snapshots of the systems
at $T=1425$ and $1500$ K. The diamondlike particles are shown as red balls and the particles with disordered local surrounding are yellow.
We see that at $T=1425$ K almost all particels in the system are diamondlike. At the same time, the system at $T=1500$ K shows a two
phase behavior: a part of the system is diamonlike, while another part is disordered.

\begin{figure}
\includegraphics[width=8cm]{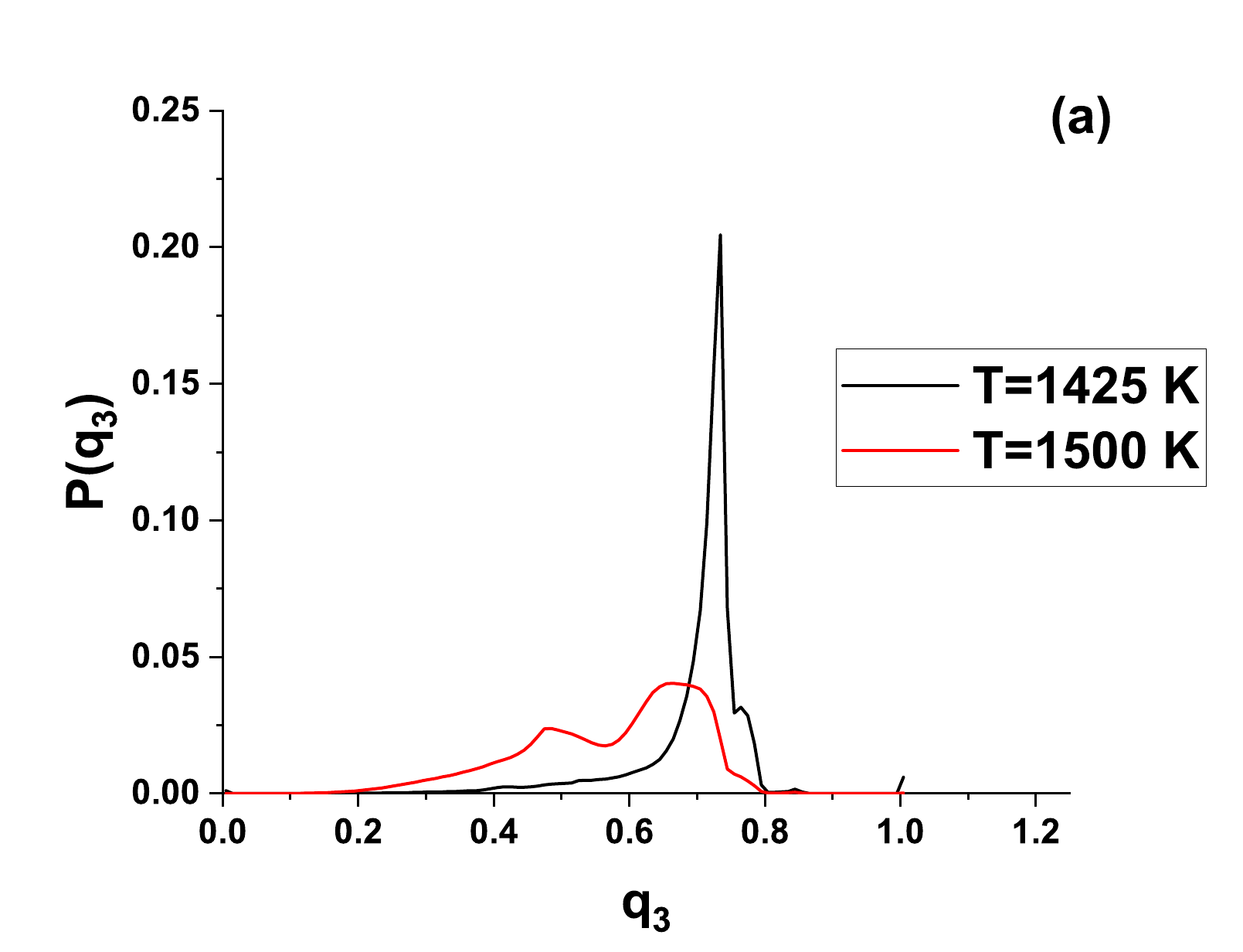}%

\includegraphics[width=8cm]{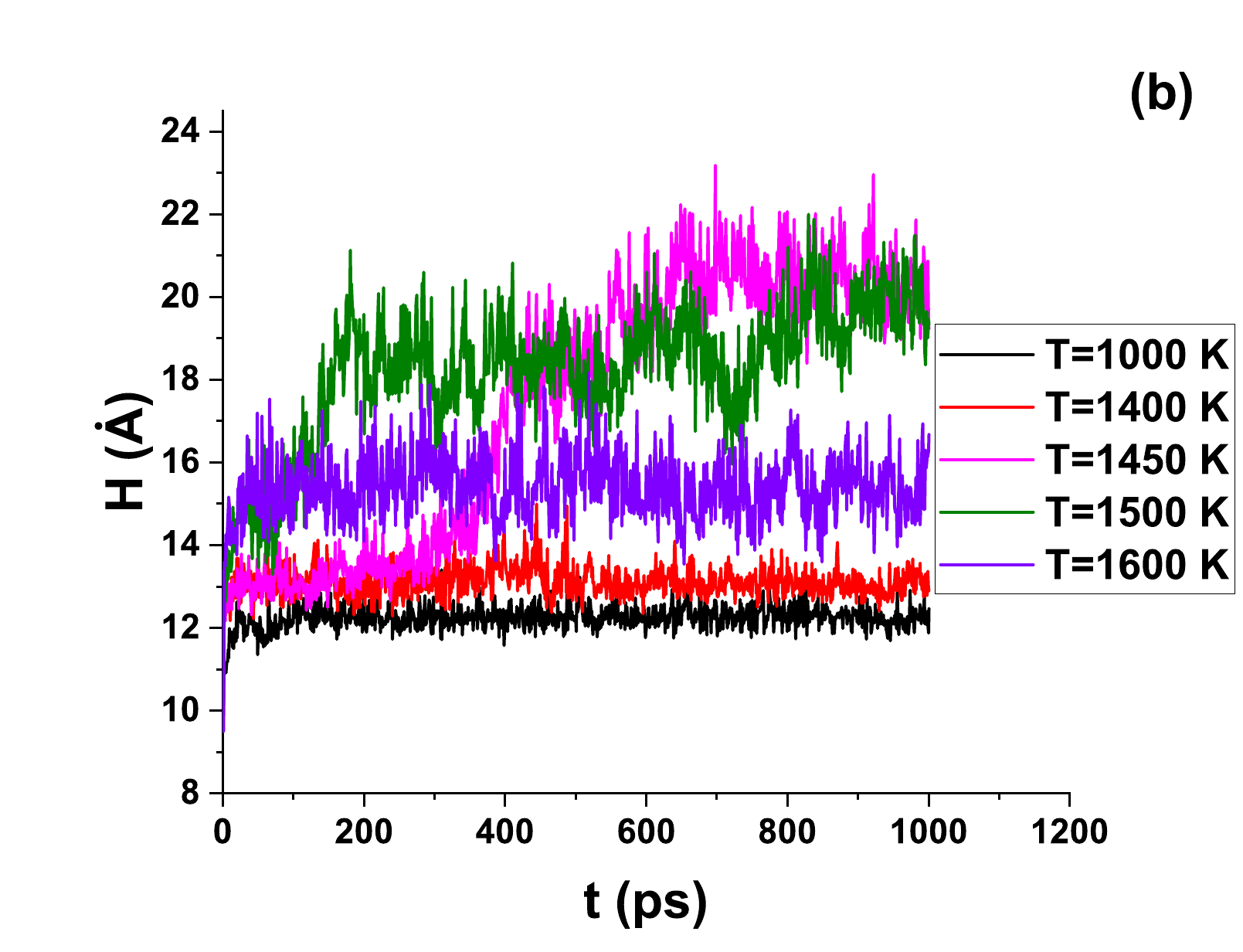}%

\caption{\label{si2-w} (a) Probability distribution of BOO $q_3$ for the system with eight layers at two temperatures. (b) The width of the sample of eight particles at different temperatures.}
\end{figure}

This conclusion is supported by calculation of BOO $q_3$. Figure \ref{si2-w} (a) shows a probability distribution of $q_3$ at $T=1425$ and $1500$ K. It is seen that
at the former temperature the probability distribution shows a single sharp peak at $q_3=0.735$ which is very close to the value in a perfect diamond crystal, while
at $T=1500$ K a two-peak distribution is observed. This is consistent with the presence of two phases in the system.

Figure \ref{si2-w} (b) shows the width of the sample at different temperatures. One can see that at all studied temperatures the system remains condensed, i.e. it does not
occupy the whole volume of the box. This is again an effect of finite time of the simulation, since the system is under the pressure below the triple point one.

\subsection{Twelve layers system}

\begin{figure}
\includegraphics[width=8cm]{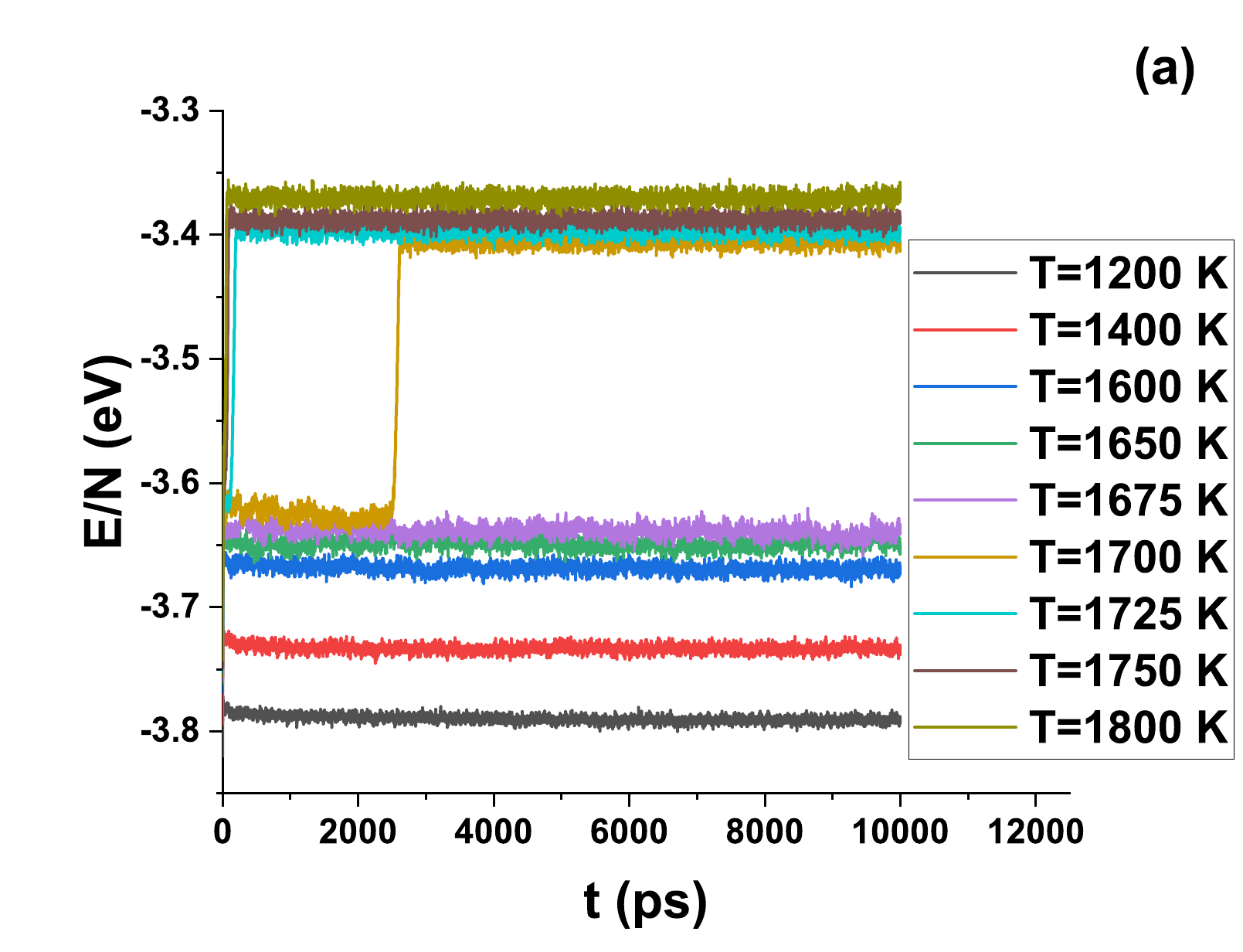}%

\includegraphics[width=8cm]{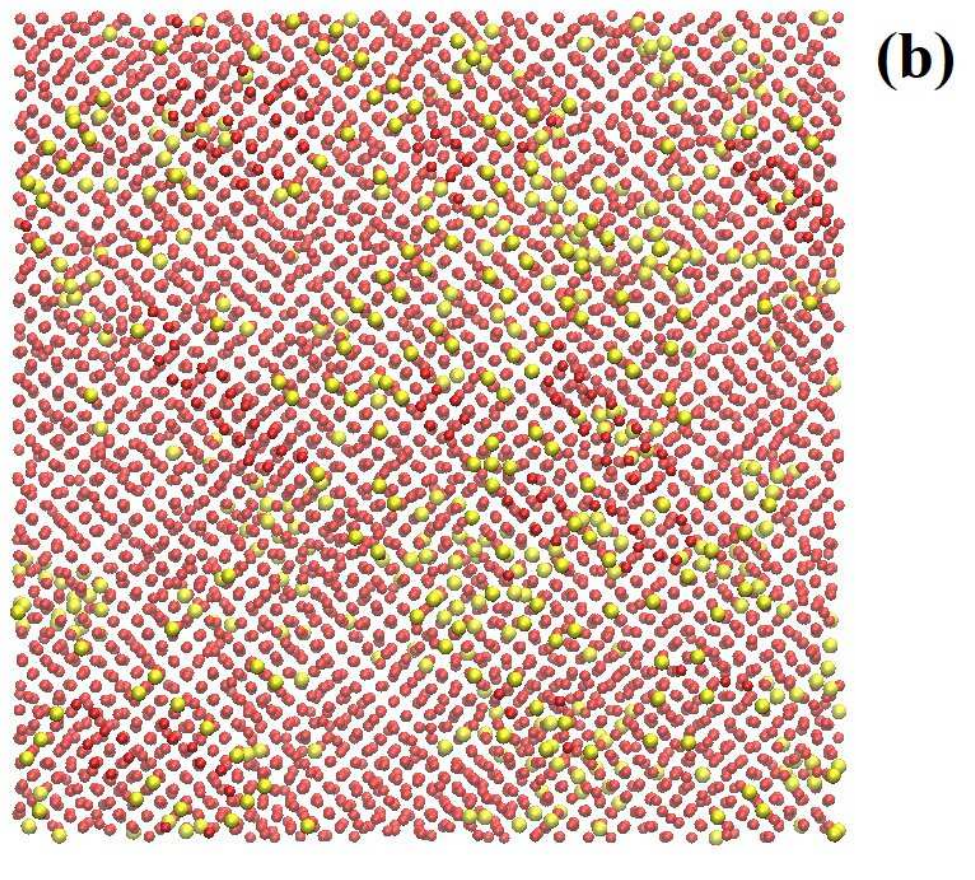}%

\includegraphics[width=8cm]{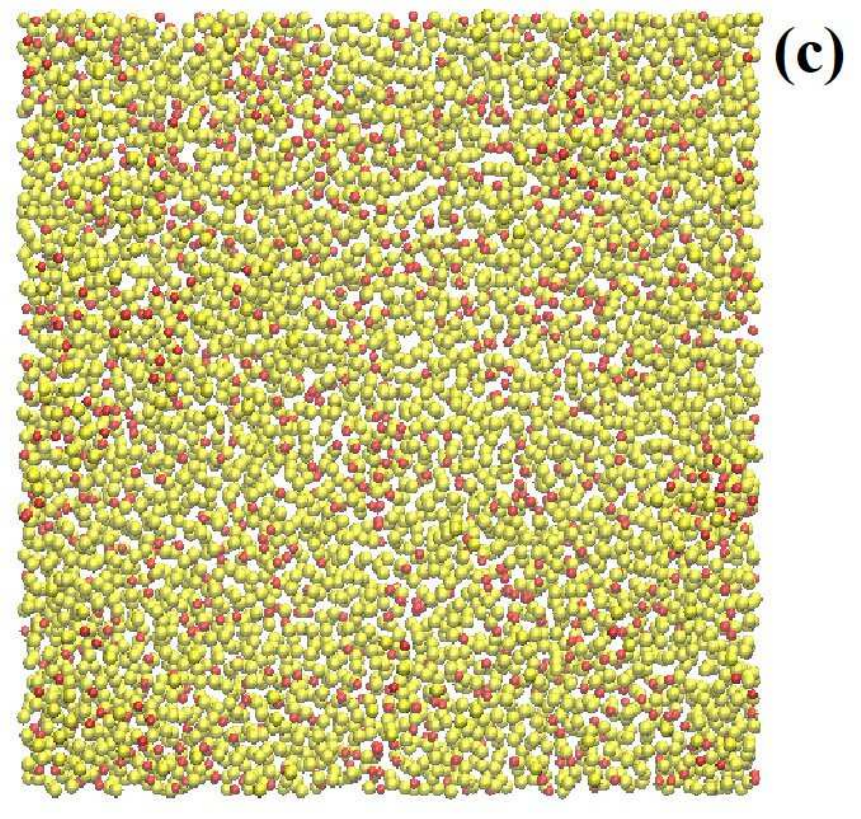}%

\caption{\label{si3} (a) The time dependence of the energy per particle of twelve layer system at different temperatures. (b) A snapshot of the system at $T=1625$ K. (c) A snapshot
of the system at $T=1700$ K. Red balls are diamondlike particles, yellow ones are the particles with disordered local surrounding (see the methons section).}
\end{figure}

Next we consider a system with twelve layers of silicon atoms. Figure \ref{si3} (a) shows a time dependence of the energy per particle of this system for a set of temperatures.
In this case an abrupt change of the energy takes place at $T=1700$ K, while at $T=1675$ K the energy fluctuates about an average value during the whole simulation. Figures \ref{si3}
(b) and (c) show snapshots of this system at $T=1675$ and $1700$ K respectively. It is seen from these snapshots, that the system is in the crystalline state at the former temperature,
but in the disordered one in the later.

\begin{figure}
\includegraphics[width=8cm]{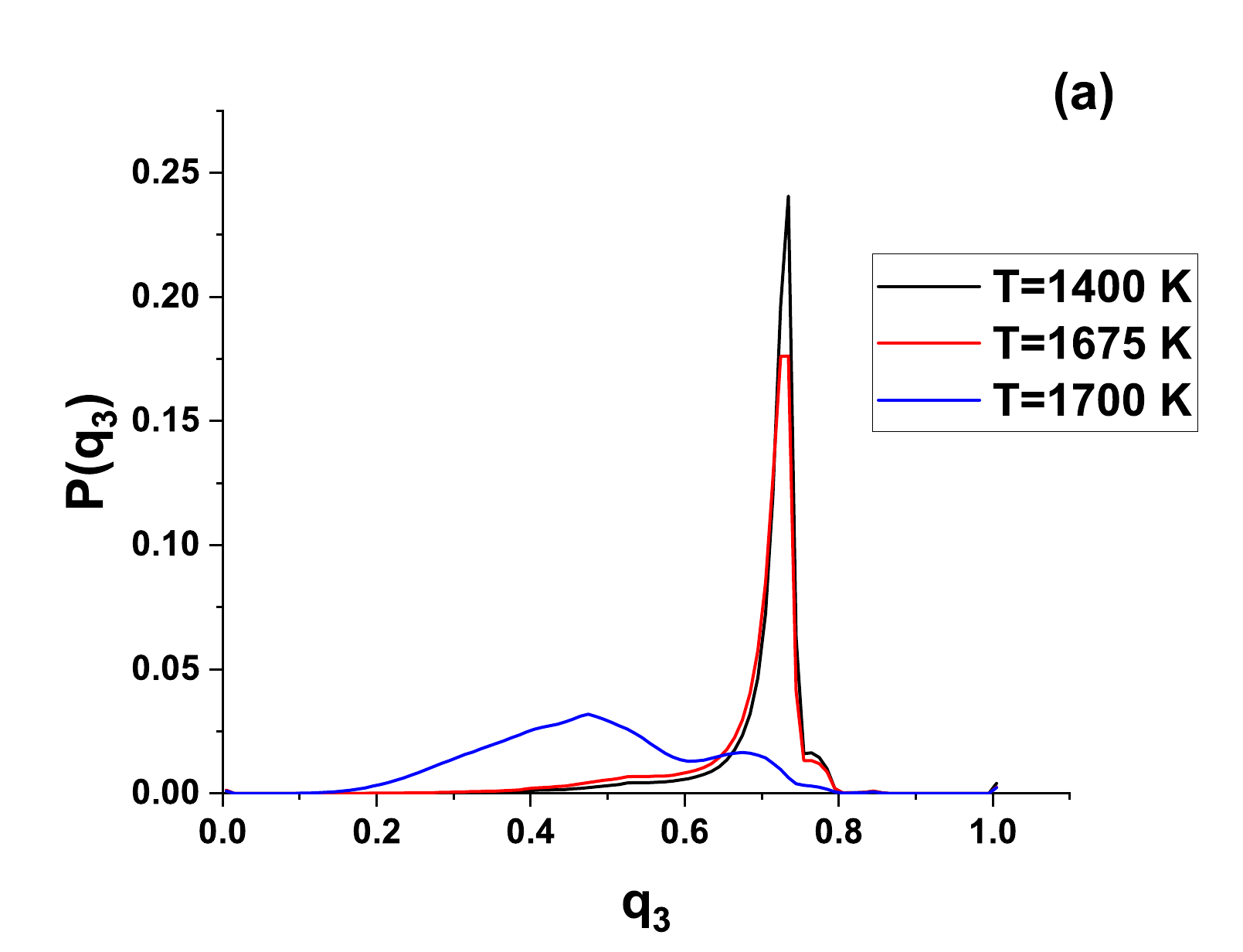}%

\includegraphics[width=8cm]{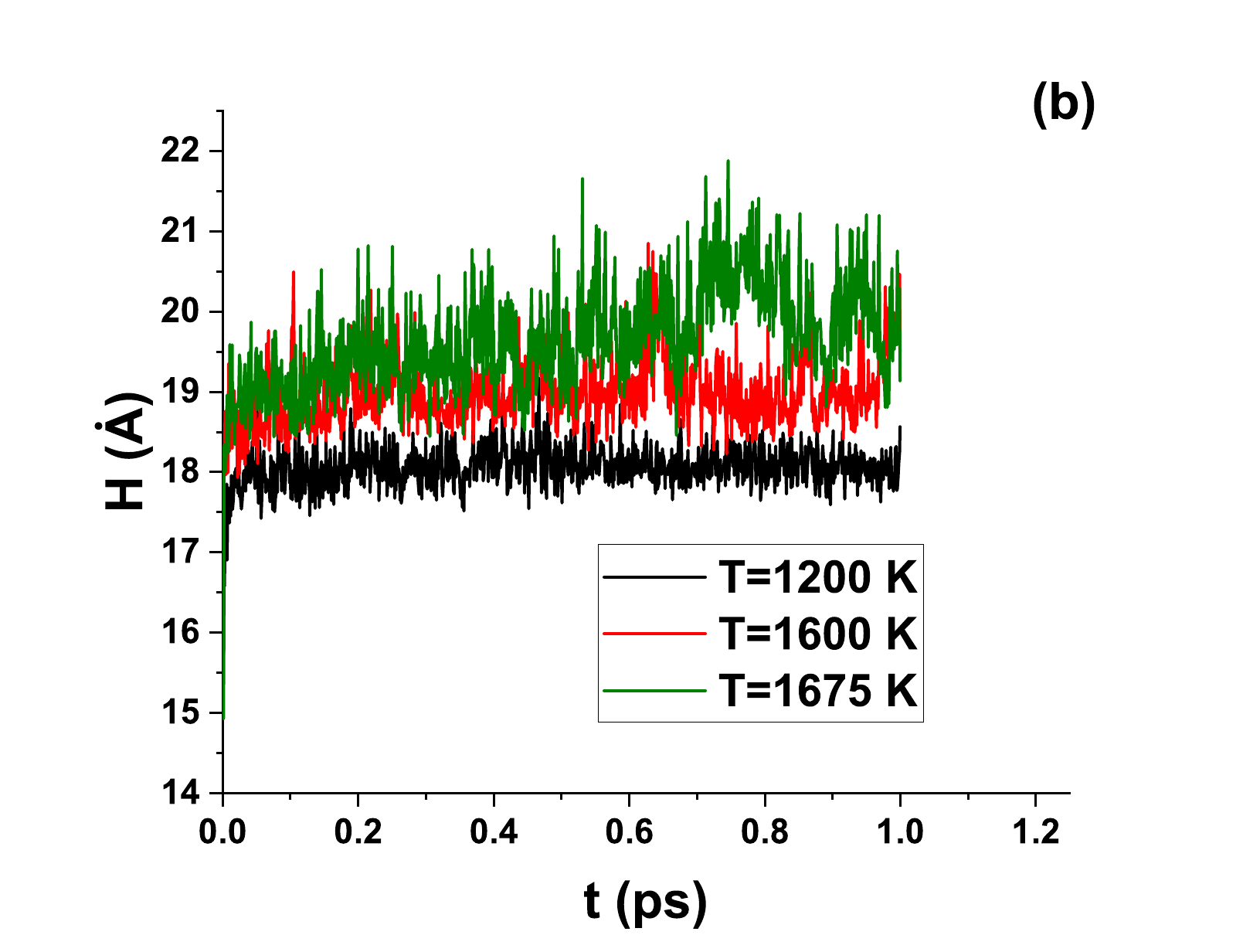}%

\caption{\label{si3-w} (a) Probability distribution of BOO $q_3$ for the system with twelve layers at two temperatures. (b) The width of the sample of twelve particles at different temperatures.}
\end{figure}

Figure \ref{si3-w} (a) shows a distribution of BOO $q_3$ at several temperatures. This plot confirms the conclusion that the crystalline structure breaks at $1700$ K. Like in the
previous cases, the system remains condensed due to finite time of the simulations, which is seen from the height of the system at different temperatures (Fig. \ref{si3-w})

\subsection{Sixteen layer system}

\begin{figure}
\includegraphics[width=8cm]{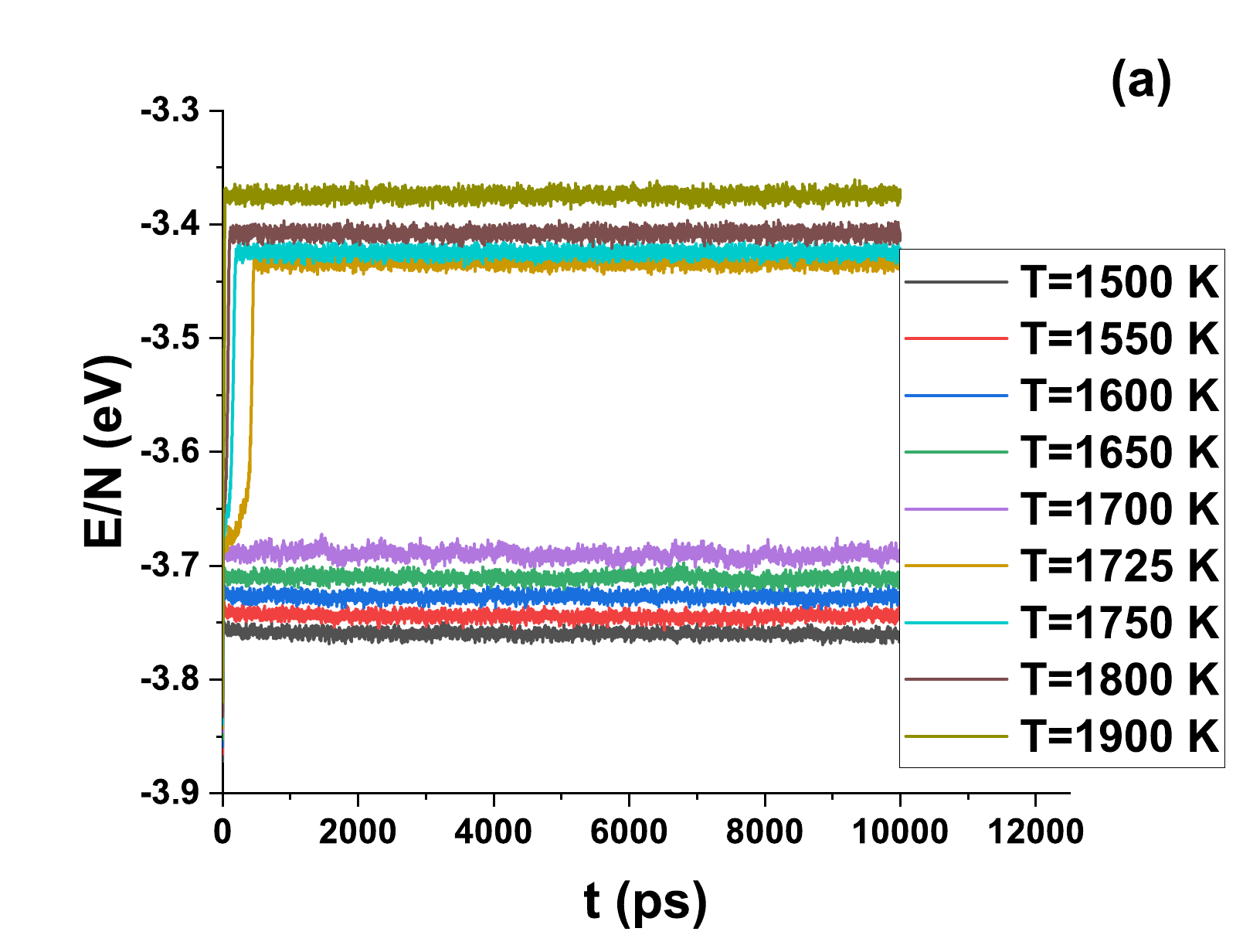}%

\includegraphics[width=8cm]{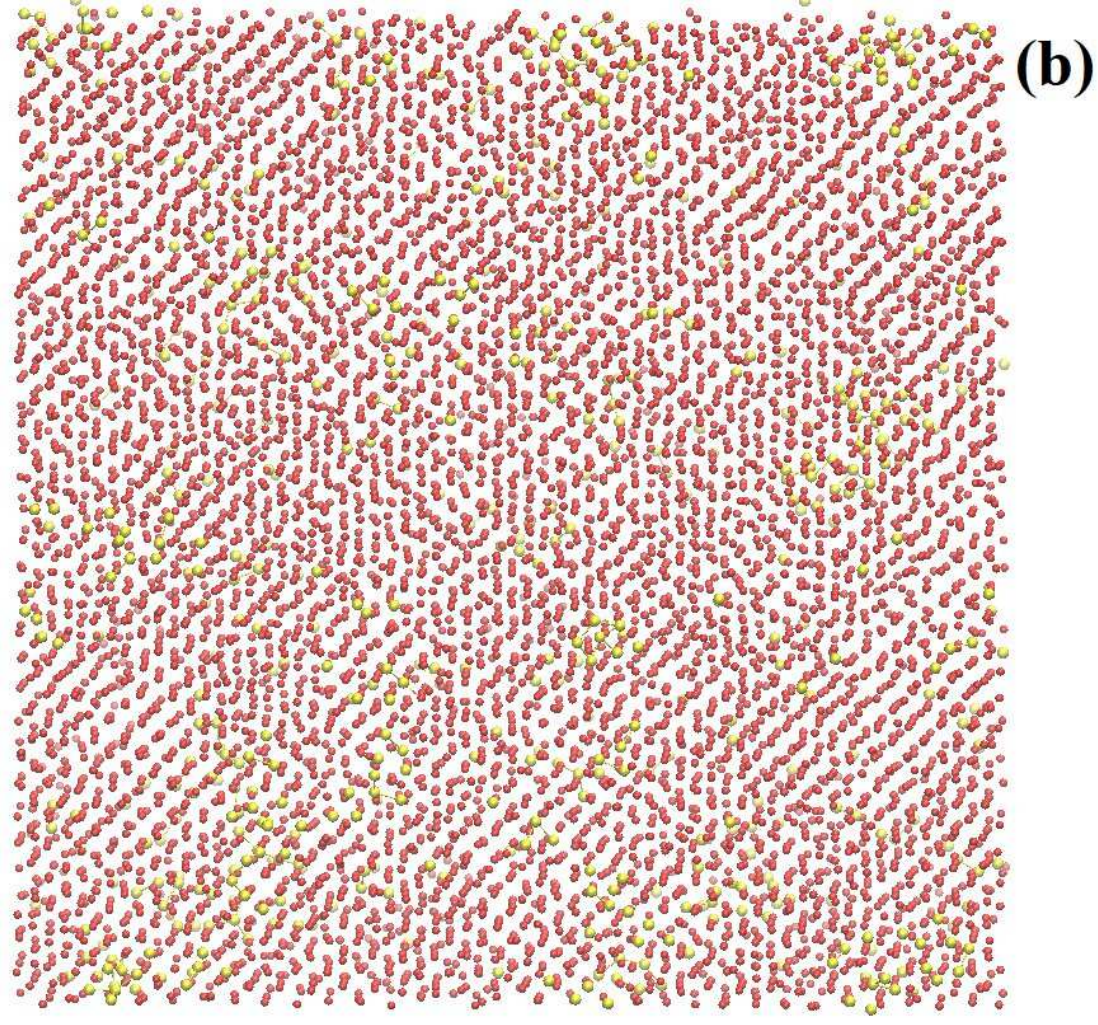}%

\includegraphics[width=8cm]{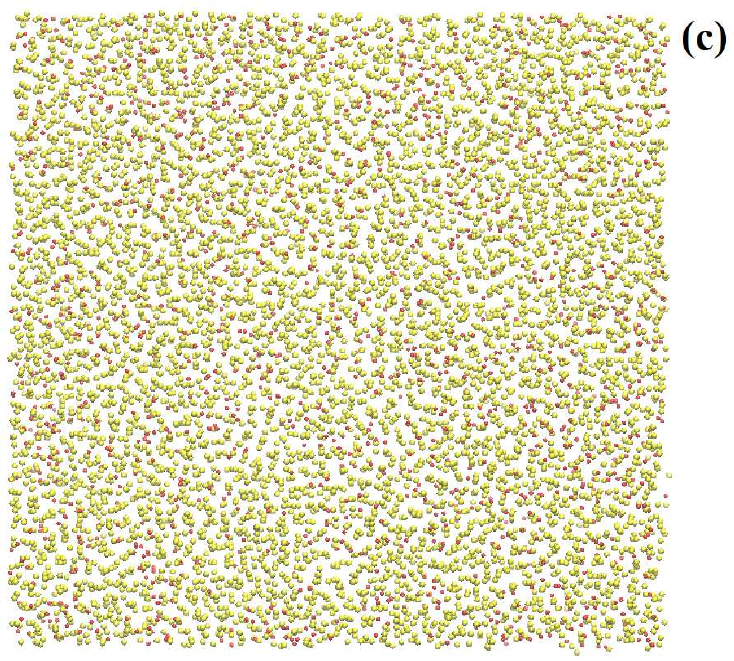}%

\caption{\label{si4} (a) Time dependence of the energy per particle of sixteen layer system at different temperatures. (b) A snapshot of the system at $T=1700$ K. (c) A snapshot
of the system at $T=1725$ K. Red balls are diamondlike particles, which yellow ones are the particles with disordered local surrounding (see the methons section).}
\end{figure}

In the case of the system with sixteen layers the crystalline order preserves at the temperatures up to $T=1700$ K, but breaks down at $T=1725$ K (Fig. \ref{si4}). Panels
(b) and (c) of Fig. \ref{si4} show snapshots of the system at these temperatures which confirm this statement.

\begin{figure}
\includegraphics[width=8cm]{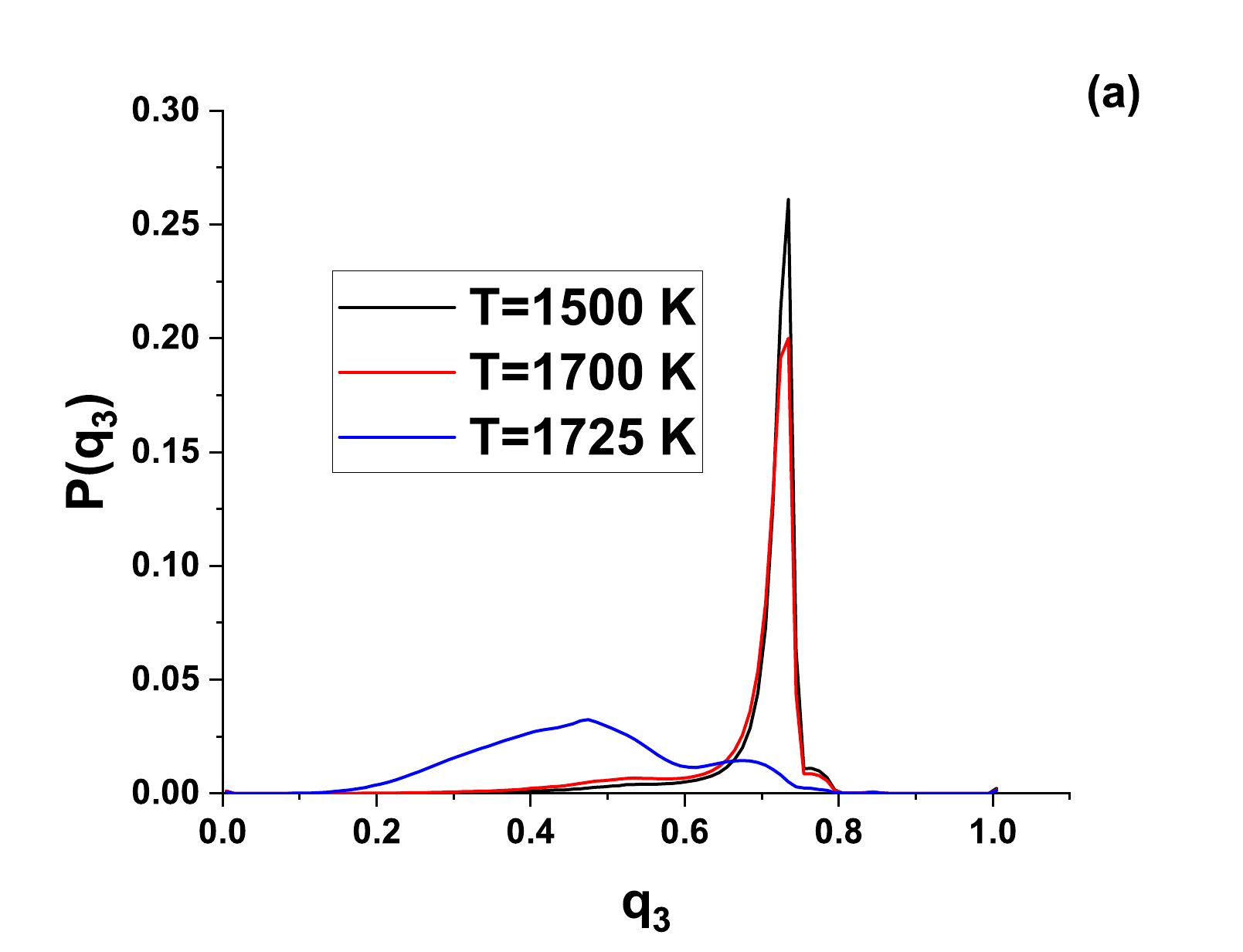}%

\includegraphics[width=8cm]{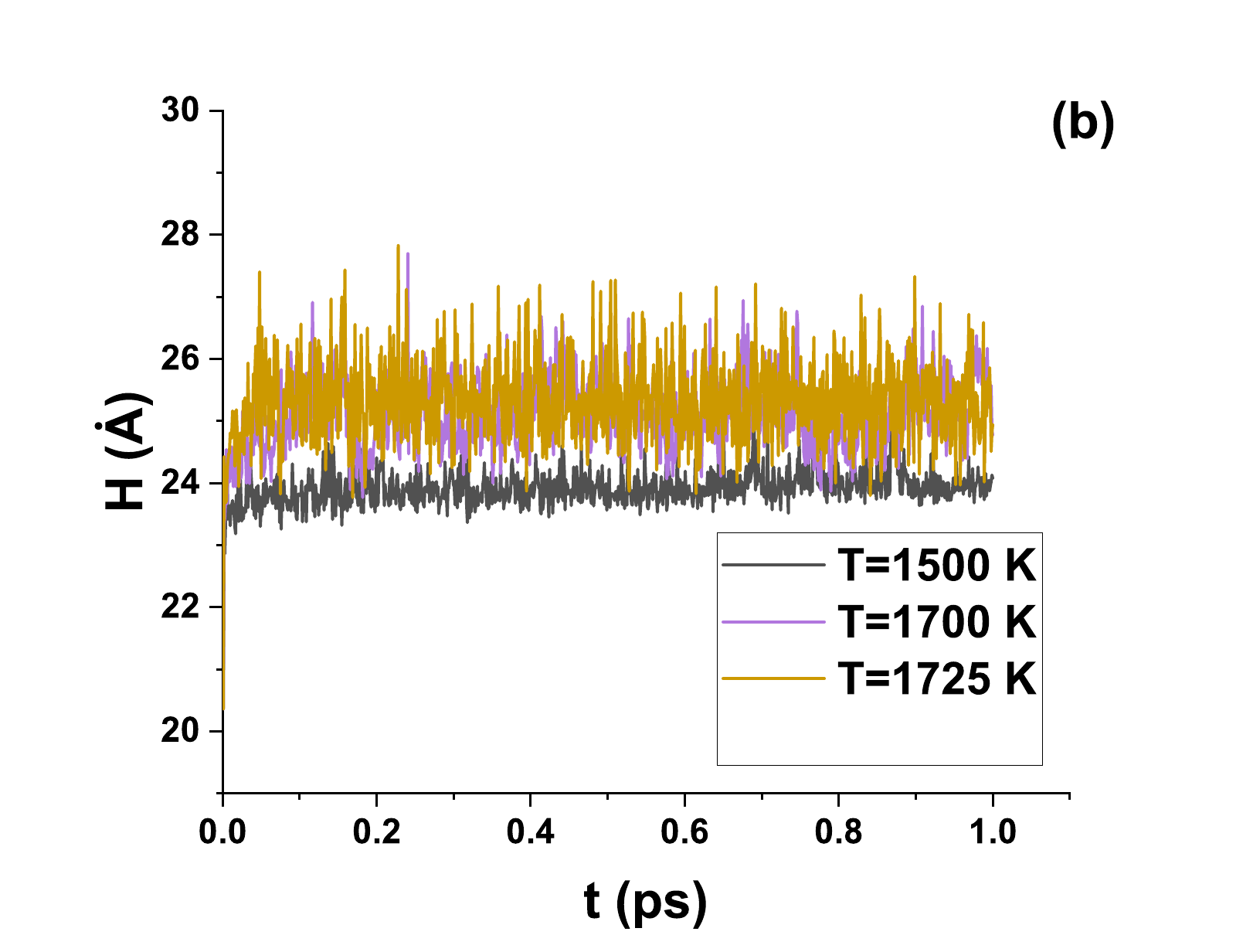}%

\caption{\label{si4-w} (a) Probability distribution of BOO $q_3$ for the system with sixteen layers at two temperatures. (b) The width of the sample of sixteen particles at different temperatures.}
\end{figure}

Unlike in the previous cases, the system looses the crystalline order almost completely at $T=1725$ K, which is seen from the probability distribution
of BOO (Fig. \ref{si4-w} (a)): the crystalline peak at this temperature is very small. At the same time the system remains condensed and its width does not
change upon breaking of the structure: the width of the system below the breaking point and above it is appoximately equal (Fig. \ref{si4-w} (b)).

\subsection{Twenty layer system}

\begin{figure}
\includegraphics[width=8cm]{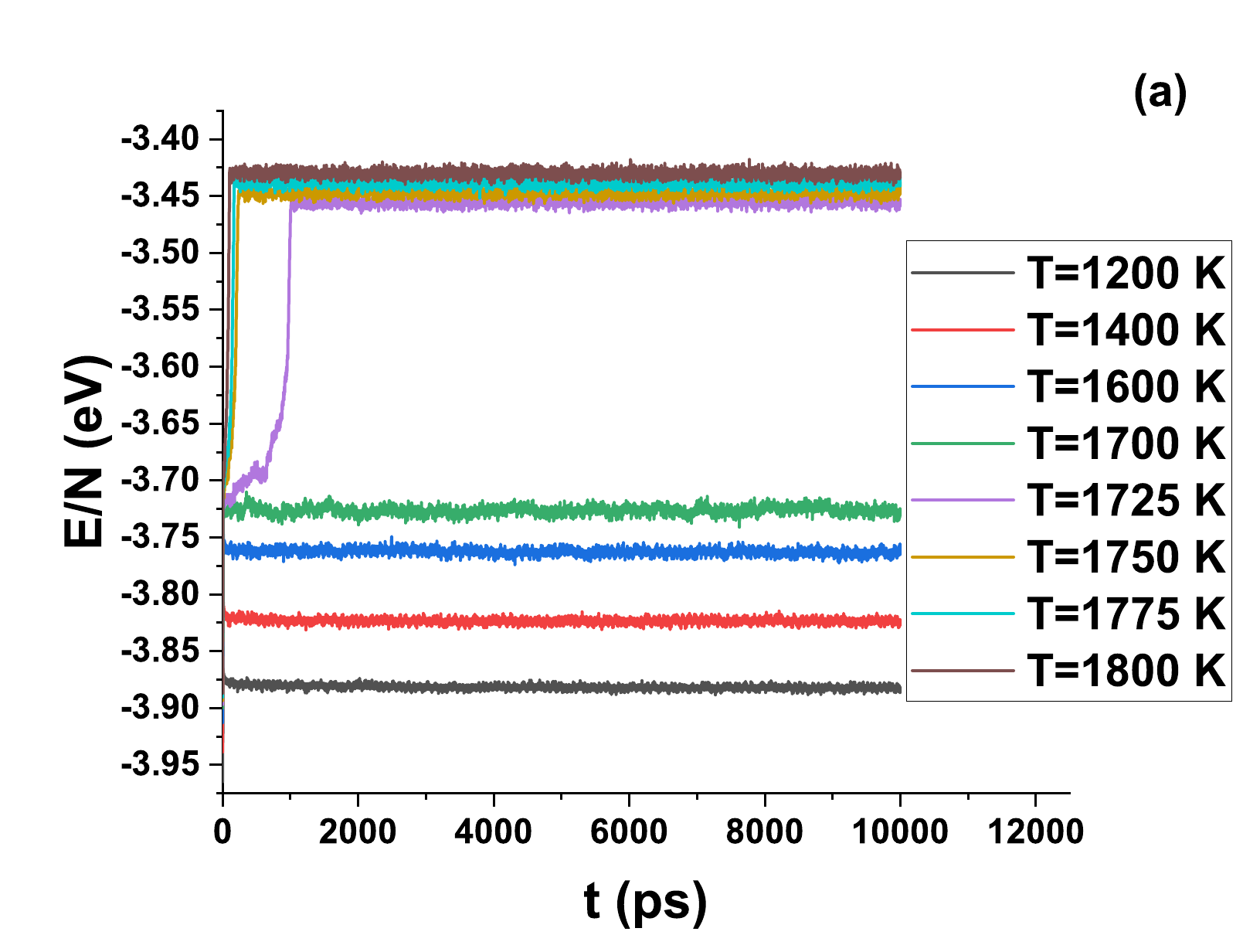}%

\includegraphics[width=8cm]{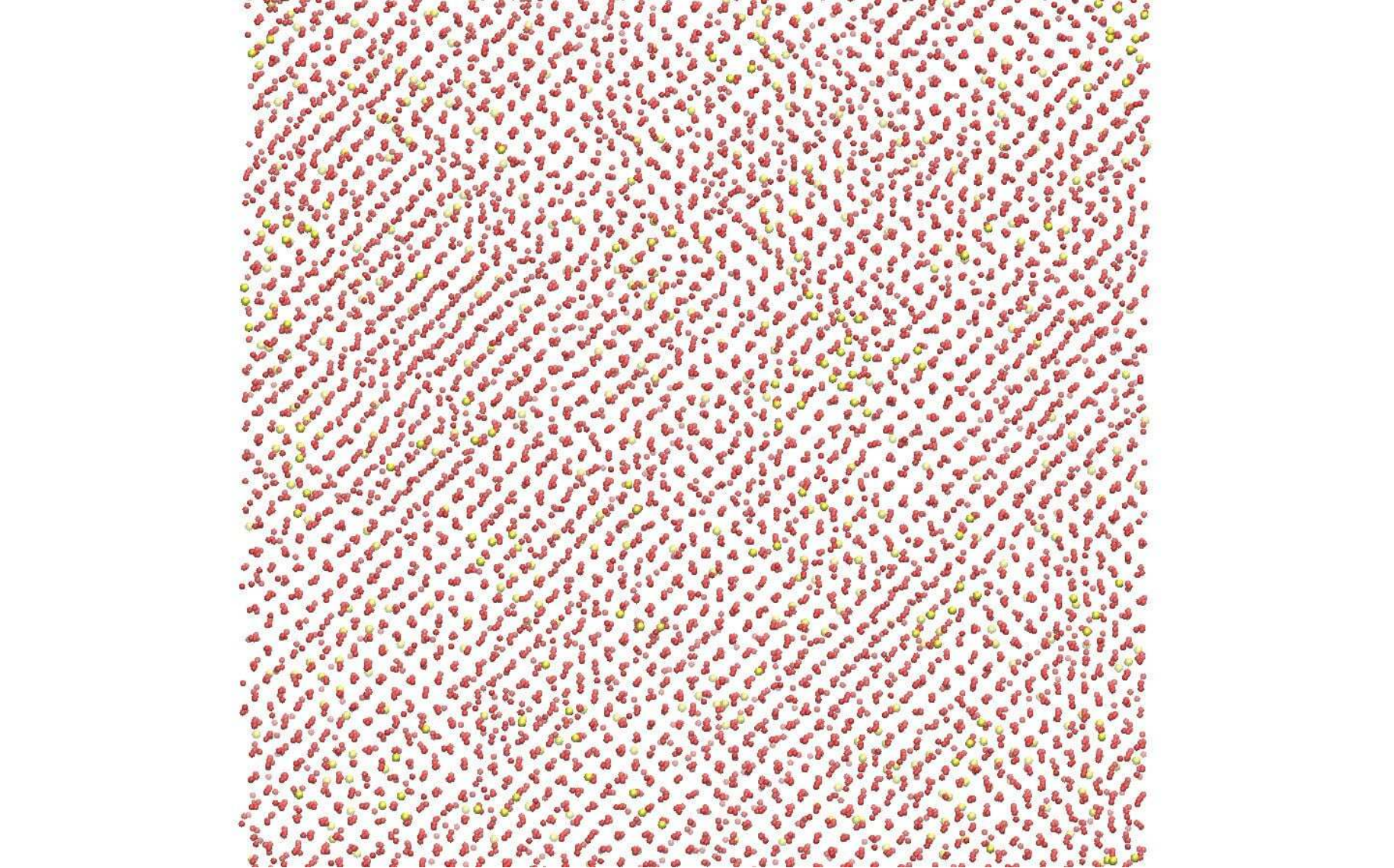}%

\includegraphics[width=8cm]{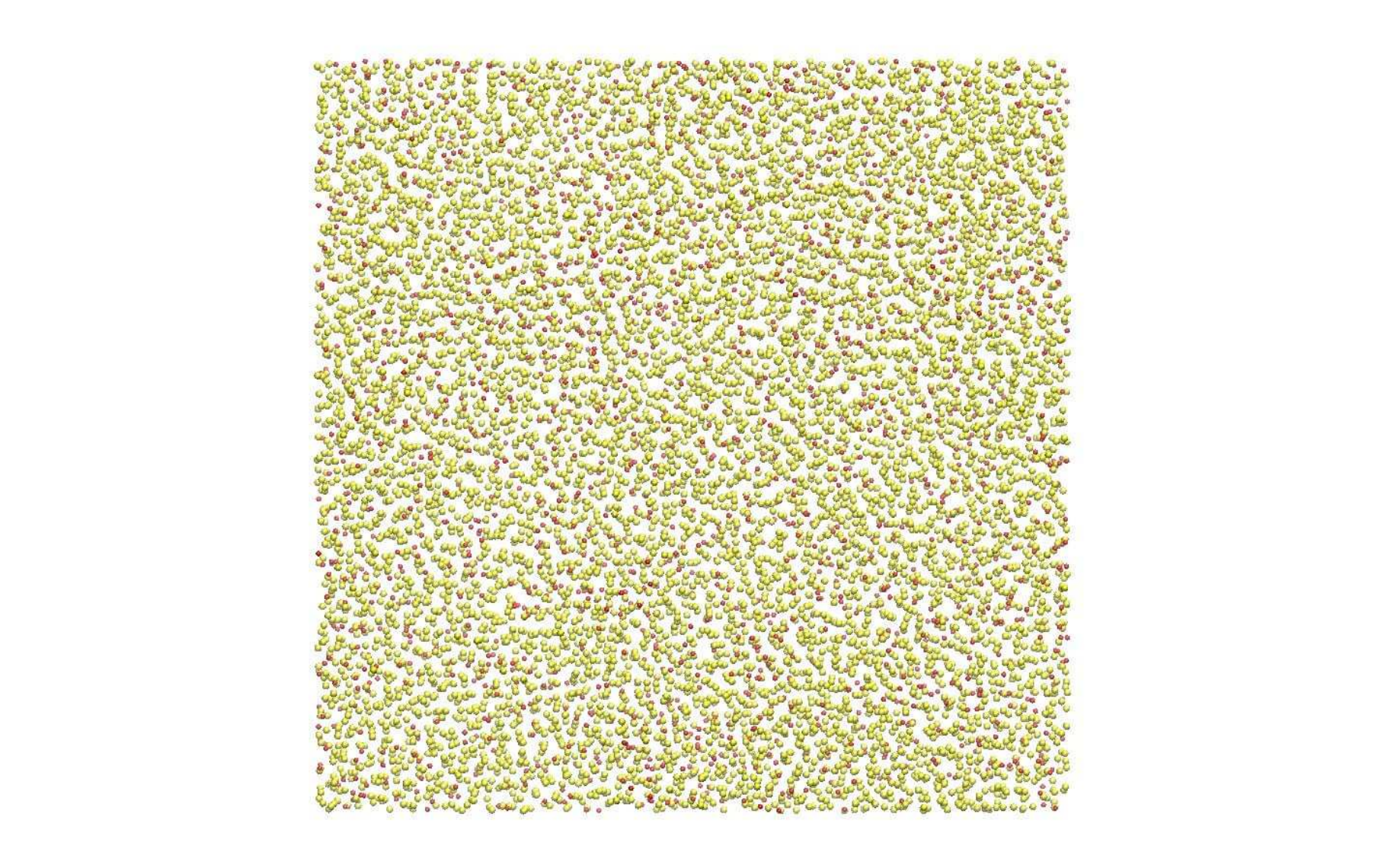}%

\caption{\label{si5} (a) Time dependence of the energy per particle of twenty layer system at different temperatures. (b) A snapshot of the system at $T=1700$ K. (c) A snapshot
of the system at $T=1725$ K. Red balls are diamondlike particles, which yellow ones are the particles with disordered local surrounding (see the methons section).}
\end{figure}

In the case of twenty layer system the results coincide with the previous case: the system remains crystalline at $T=1700$ K, but the crystal structure breaks down at $T=1725$ K.
Figure \ref{si5} (a) shows the evolution of potential energy per atom, where a suddent jump is observed at $T=1725$ K. Panels (b) and (c) of the same figure demonstrate
snapshots of the system at $1700$ and $1725$ K respectively. One can see that the system has diamond structure at the former temperature, but it is disordered at the later one.

\begin{figure}
\includegraphics[width=8cm]{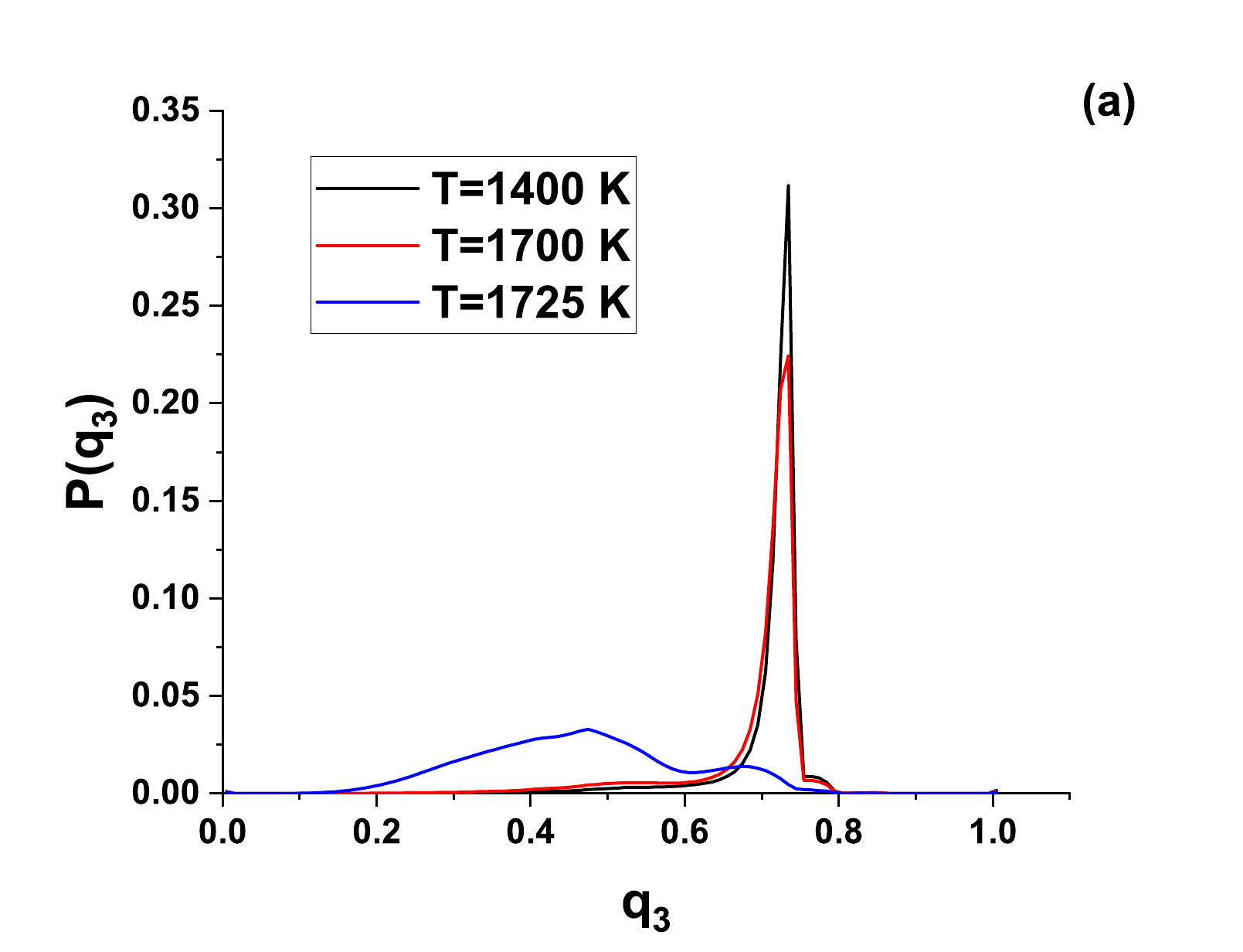}%

\includegraphics[width=8cm]{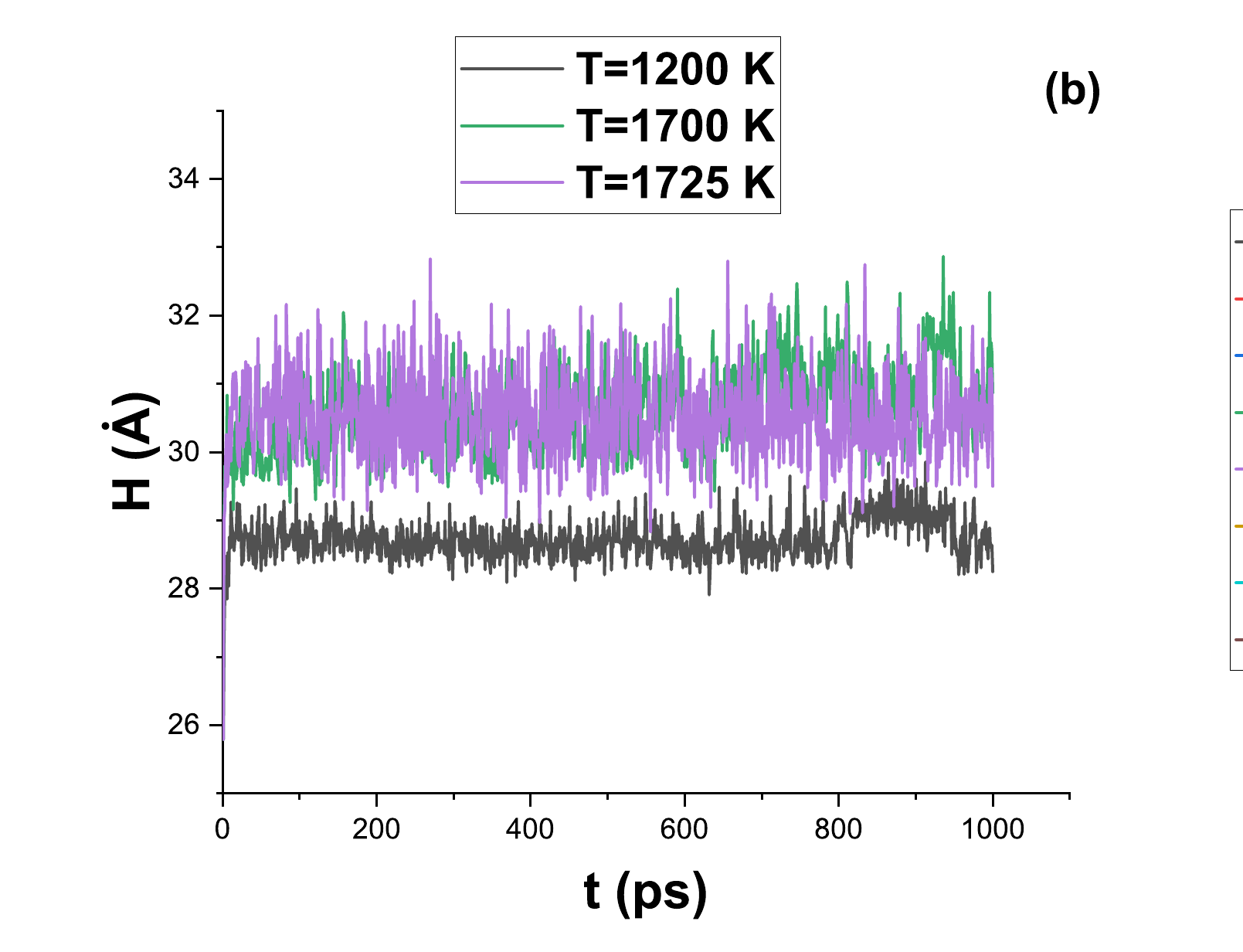}%

\caption{\label{si5-w} (a) Probability distribution of BOO $q_3$ for the system with twenty layers at two temperatures. (b) The width of the sample of twenty particles at different temperatures.}
\end{figure}

Like in the case of sixteen layer system, the probability distribution of the BOO $q_3$ demonstrates a two-peak structure after the crystal structure collapse (Fig. \ref{si5-w} (a)). The system
remains condensed, which is once again an effect of the finite simulation time (Fig. \ref{si5-w} (b)).

\subsection{Bulk system}

\begin{figure}
\includegraphics[width=8cm]{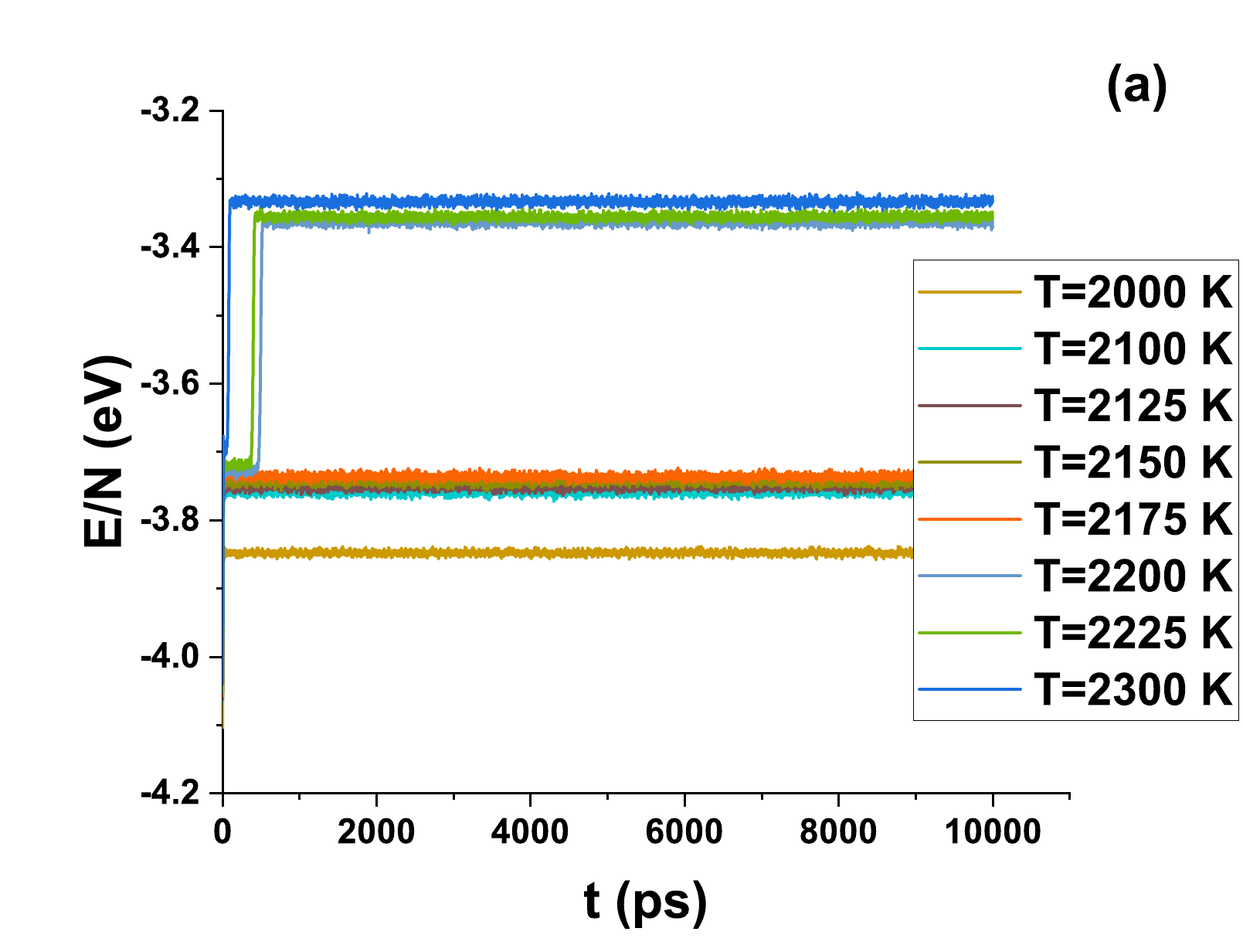}%

\includegraphics[width=8cm]{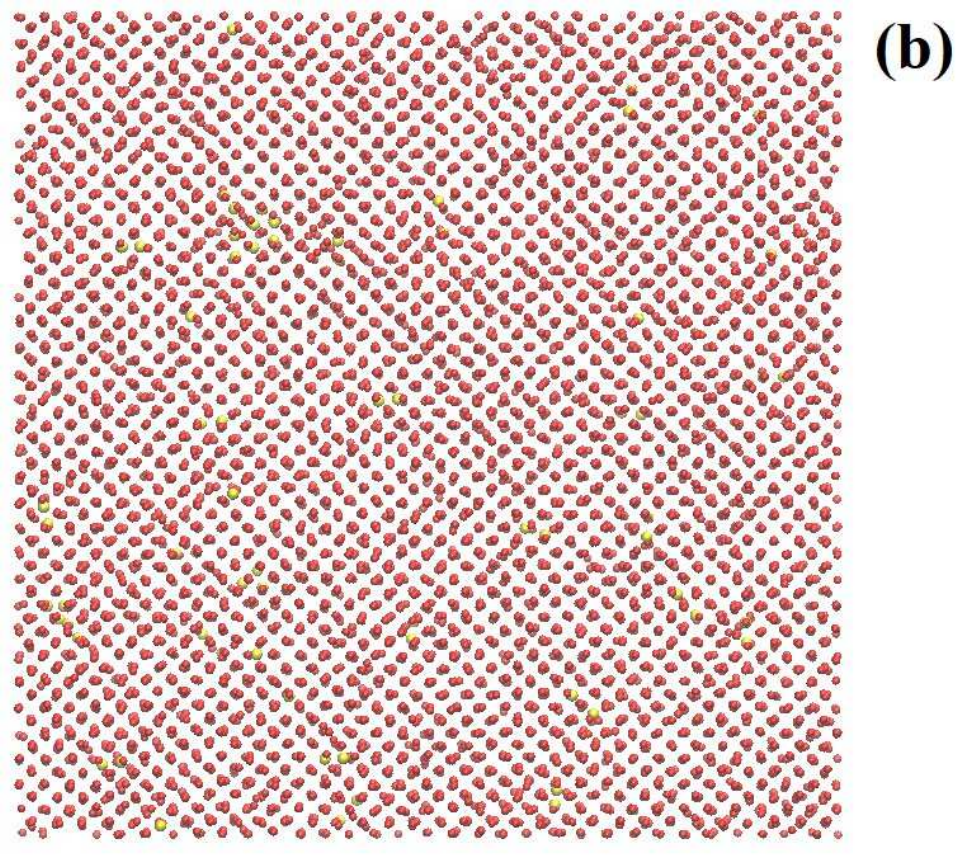}%

\includegraphics[width=8cm]{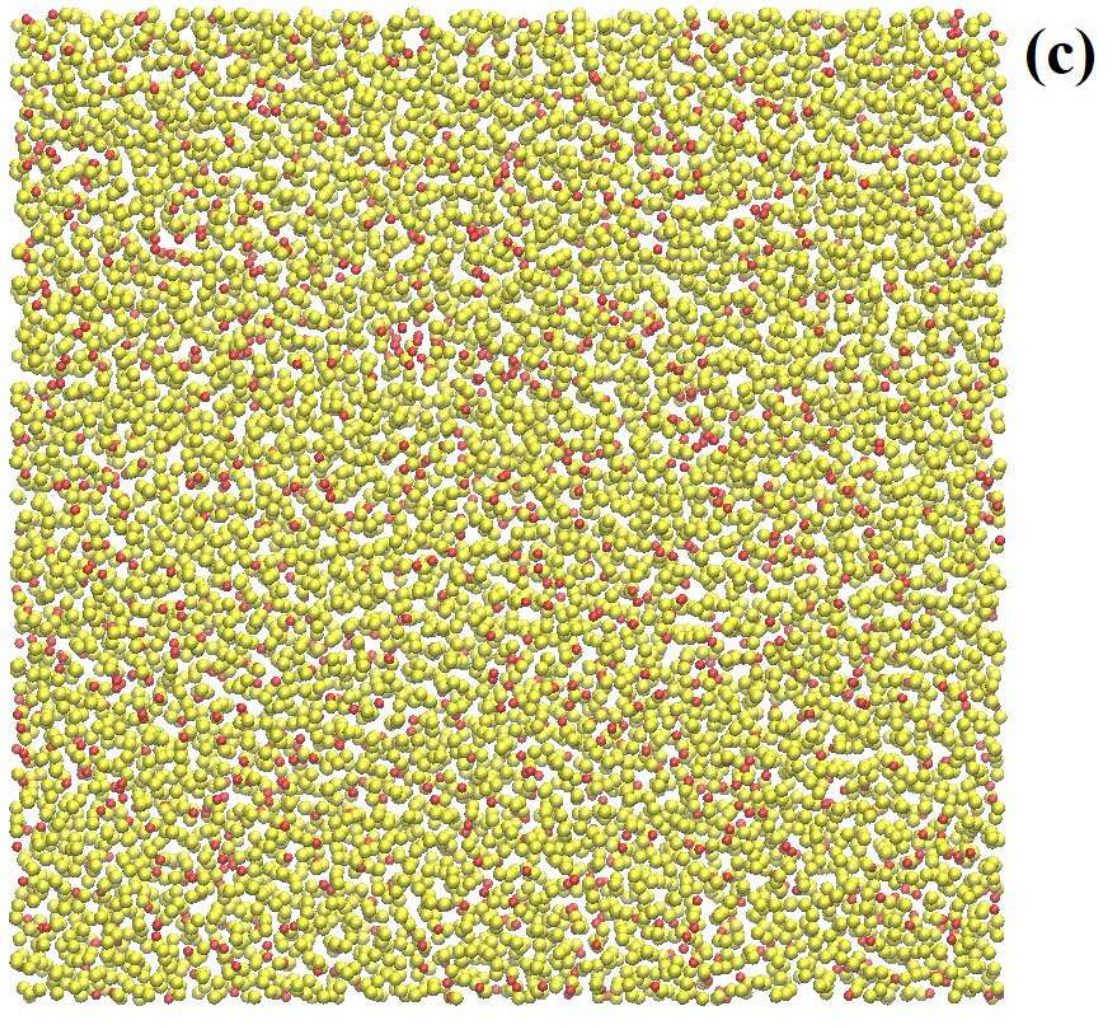}%

\caption{\label{bulk} (a) Time dependence of the energy per particle of twenty layer system at different temperatures. (b) A snapshot of the system at $T=2175$ K. (c) A snapshot
of the system at $T=2200$ K. Red balls are diamondlike particles, which yellow ones are the particles with disordered local surrounding (see the methons section).}
\end{figure}

Finally we compare the results for the films of different thikness with the results for the bulk silicon. Figure \ref{bulk} (a) shows the evolution of energy for the sample
at different temperatures. It is seen that the energy remains constant for the temperatures up to $2175$ K, but experiences a jump at $T=2200$ K. Panels
(b) and (c) of the same figure demonstrate snapshots of the system at these two temperatures. One can see that it is diamond-like at the former temperature,
but liquid-like at the later. The pressure of the system is negative $P=-27.5$ kbar, therefore sublimation takes place in this case. This spontaneous sublimation takes place at $T_s=2200$ K.
This temperature strongly exceeds the experimental melting point of silicon $T_m^{exp}=1683$ K.
However, it is well known that simple 'heat until it melts' method usually overestimates the melting point. Apparentely, the same effect should be expected in the case of sublimation.
For this reason we believe that the results are reliable.

\begin{figure}
\includegraphics[width=8cm]{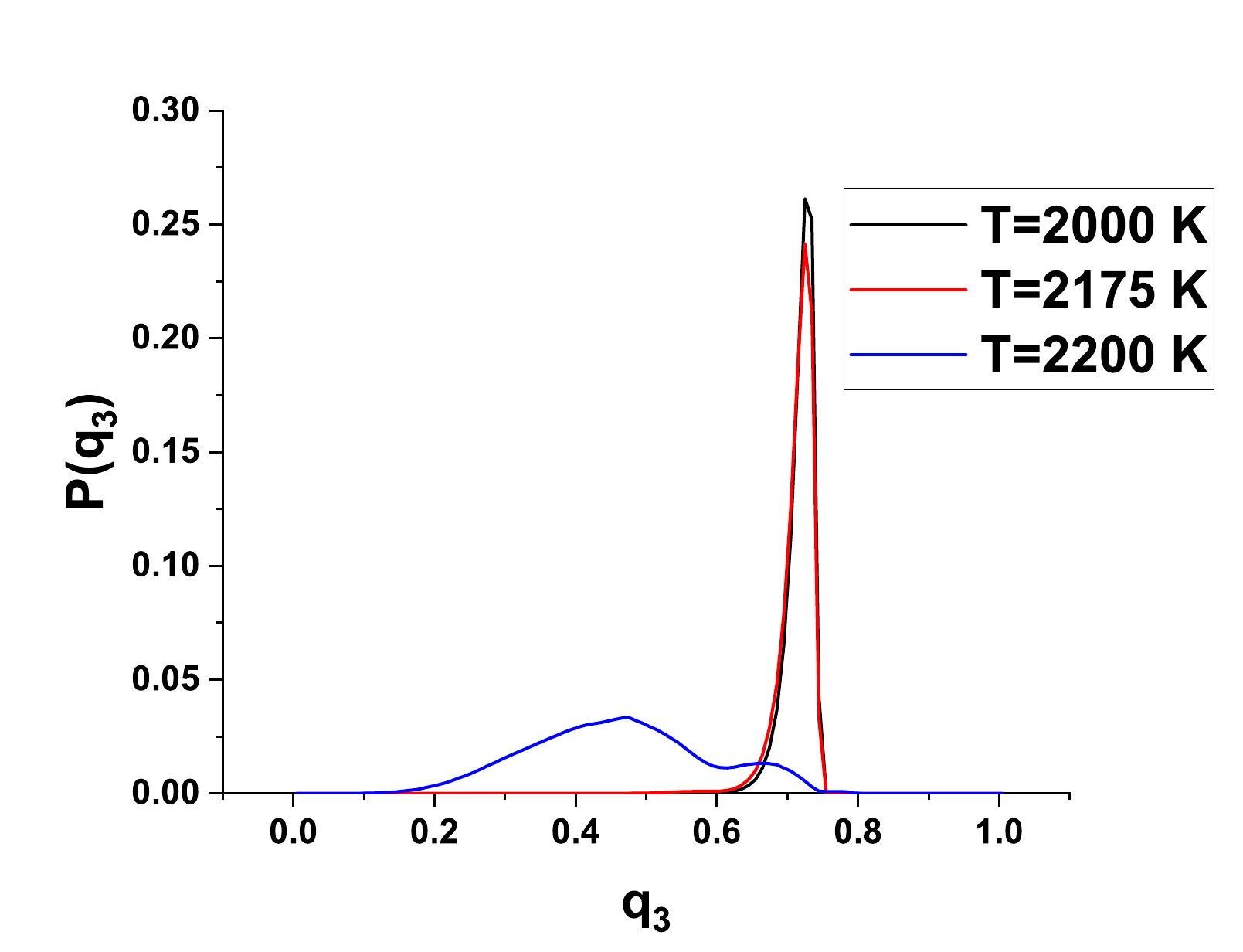}%

\caption{\label{b-q3} Probability distribution of BOO $q_3$ for the bulk system.}
\end{figure}

The probability distribution of BOO $q_3$ of the bulk silicon at three different temperatures is shown in Fig. \ref{b-q3}. The overall picture looks similar to the cases above:
it demonstrates a sharp peak arond the $q_3^{id}$ in the crystalline phase up to the melting point, and two peaks structure above it. The square of the peak
at crystalline value of $q_3$ is very small, which means that just a few particles belong to some crystal-like clusters, which is expected in a liquid
phase in the vicinity of the melting point.

\bigskip

In the present study we simulate the sublimation process of silicene, films of silicon of different width and bulk silicon. A comparison of sublimation
temperature of graphene and graphite, obtained in non-equilibrium molecular dynamics study, is given in Ref. \cite{or}.
The authors come to the conclusion that the sublimation (erroneously called 'melting' in this paper) of graphene takes place at much higher temperature
than the one of graphite. They explain it by the fact that at high temperature atoms of graphite can make bonds between different layers due to very strong
thermal fluctuations, while the atoms of graphene cannot and sublimation of graphene is proliferated by Stone-Wales defects \cite{sw-def}.

Our result allow to conclude that sublimation of silicene and silicon films goes in a different from graphite way. The in-layer bonds of carbon atoms in graphite
is very strong, and the structure breaks down due to formation of more bonds with the atoms from other layers. In the case of silicon the bond energy is  smaller
and as a result the crystal structure collapses due to bond breaking and evaporation of the particles at the interface of the fiml. As a consequence, thin films sublimate
at lower temperatures. Although, when the number of silicon layers reach $16$ the sublimation temperature comes to a saturated value of $1725$ K, which is
very close to the experimental melting point of silicon.

\begin{table}\label{t2}
\begin{tabular}{ l c r }
  System & Sublimation temperature (K) \\
\hline
  silicene & 875  \\
\hline
  8 layers  & 1450 \\
\hline
 12 layers & 1700 \\
\hline
 16 layers & 1725 \\
\hline
 20 layers & 1725 \\
\hline
 bulk  & 2200 \\
\hline
\end{tabular}
\caption{Number of atoms in systems under consideration.}
\end{table}

\section{Conclusions}

Sublimation of silicon films of different thiknes - from one layer (silicene) up to twenty layers is studied by means of molecular dynamics simulation. The results
are compared to the ones of the bulk sample. It is shown that the sublimation temperature increases with the film thinkes up to the saturation value of $1725$ K for the
films with more than $16$ atomic layers (see Table \ref{t2}). These results are consistent with the usual view on the melting of a crystal, where melting starts from the surface. However,
they are in contradiction to the mechanisms of graphite melting and sublimation, which demonstrate the difference between the high temperature behavior
of silicon and carbon.

This work was carried out using computing resources of the federal
collective usage center "Complex for simulation and data
processing for mega-science facilities" at NRC "Kurchatov
Institute", http://ckp.nrcki.ru, and supercomputers at Joint
Supercomputer Center of the Russian Academy of Sciences (JSCC
RAS). The work on the bulk silicon was supported by Russian Science Foundation (Grant 19-12-00111) and
the work on silicene and silicon films was supported by Russian Science Foundation (Grant  19-12-00092).

\end{document}